%% file: paper_main.tex
\documentclass[submission, Phys]{SciPost}

\usepackage{microtype}
\usepackage{subcaption}
\usepackage{multirow}
\usepackage{array}
\usepackage{enumitem}
\usepackage{float}
\usepackage{amsmath}
\usepackage{amssymb}
\usepackage{graphicx}
\usepackage{xcolor}
\usepackage{booktabs}
\usepackage{nicefrac}
\usepackage{orcidlink}
\usepackage[symbol]{footmisc}

\pdfoutput=1
\newcommand{\ParT}{\texttt{ParT}}

\hypersetup{
    colorlinks=true,
    linkcolor=black,
    citecolor=blue,
    urlcolor=blue
}

\begin{document}

\begin{center}
{\Large \textbf{Dissecting Jet-Tagger Through Mechanistic Interpretability %:\\[0.3em]
%A Case Study for Top Quark Jet Tagging}
}}

\vspace{0.6em}

Saurabh Rai\orcidlink{0009-0004-1080-2370}\footnote[1]{\href{mailto:saurabhrai25@iitk.ac.in}{saurabhrai25@iitk.ac.in}}, Sanmay Ganguly\orcidlink{0000-0003-1285-9261}\footnote[2]{\href{mailto:sanmay@iitk.ac.in}{sanmay@iitk.ac.in}} \\
% \vspace{0.4em}
% {\small \href{mailto:saurabhrai25@iitk.ac.in}{saurabhrai25@iitk.ac.in},\;
% \href{mailto:sanmay@iitk.ac.in}{sanmay@iitk.ac.in}}

\vspace{0.4em}

{ Department of Physics, Indian Institute of Technology, Kanpur. \\ Uttar-Pradesh 208016, India}\\[0.2em]

\vspace{0.8em}

\today

\end{center}

% ══════════════════════════════════════════════════════════════════
\begin{abstract}
\noindent
\textbf{Mechanistic interpretability seeks to reverse engineer a trained neural network by identifying the minimal subset of internal components.
We perform a mechanistic interpretability analysis of the Particle Transformer architecture,
trained on the Top Quark Tagging reference dataset, with the goal of
identifying the computational circuit responsible for jet classification
and characterizing the physical content of its internal representations.
Combining zero ablation, path patching with
two complementary on-manifold corruption strategies and linear probing of
the residual stream, we identify a sparse six-head circuit that recovers
the great majority of the full model performance while admitting a clean
source-relay-readout interpretation. In this circuit, a single early layer head
serves as the primary causal source, a cluster of middle-layer heads acts
as relays selectively attending to hard pairwise substructure and a
single late-layer head reads out the aggregated signal. Linear probes show that the residual stream is preferentially aligned
with the energy correlator basis over the $N$-subjettiness basis,
with the advantage surviving residualization against jet mass.
Within the energy correlator basis, the model preferentially encodes
2-prong substructure observables over the 3-prong observables that
are the formally correct targets for top tagging, indicating that the
network has implicitly factorized the 3-prong classification task
into the more accessible sub-problem of identifying the hadronic
$W$-boson decay.
A per-layer trained probe further reveals that the apparent single step
commitment of the model to a classification decision in the first class
attention block is in fact a basis rotation, with the discriminating signal
already saturating in the particle attention stack. These results demonstrate
that mechanistic interpretability methods developed for natural language
models can be used for jet physics classifiers and indicate that gradient descent may
rediscover physically meaningful aspects of jet tagging without
supervision.}
\end{abstract}

% \vspace{0.5em}
% \noindent
% \begin{tabular}{@{}p{0.18\textwidth}p{0.78\textwidth}@{}}
%   \textbf{Copyright:} &
%     \href{https://creativecommons.org/licenses/by/4.0/}{CC-BY 4.0}.
%     This work is licensed under a Creative Commons Attribution 4.0
%     International License. \\[4pt]
%   \textbf{Received:} & [date] \\
%   \textbf{Published:} & [date] \\[4pt]
%   \textbf{doi:} & \href{https://doi.org/10.21468/SciPostPhys.XXX}{10.21468/SciPostPhys.XXX}
% \end{tabular}

\newpage
\vspace{\baselineskip}

\noindent\textcolor{white!90!black}{\rule{\linewidth}{1pt}}

\tableofcontents

\noindent\textcolor{white!90!black}{\rule{\linewidth}{1pt}}

\vspace{\baselineskip}

%============== 1. INTRODUCTION ================= %
\input{sections/introduction}

%============== 2. EXISTING LITERATURE ================= %
\input{sections/existing_literature}

%============== 3. MECH-INTERP================= %
\input{sections/mechanistic_interpretability}

%============== 4. EXP-SETUP ================ %
\input{sections/exp_setup}

%============== 5. LOGIT-LENS ================ %
\input{sections/logitlens}

%============== 6. CIRCUIT-IDENTIFICATION ================ %
\input{sections/circuit_identification}

%============== 7. PHYSICAL-CONTENT ================ %
\input{sections/physical_content}

%============== 8. FEATURES ================ %
\input{sections/features_sec}

%============== 9. DISCUSSION ================ %
\input{sections/discussion}

%============== 10. CONCLUSION ================ %
\input{sections/conclusion}

%============== 11. ACKNOWLEDGEMENT ================ %
\input{sections/acknowledgment}

%============== APPENDIX ================= %
\input{sections/appendix}
% Flush any remaining deferred floats BEFORE the bibliography so they
% don't end up printed after the references.
%\clearpage
\bibliography{bibliography}
\end{document}

%% file: sections/introduction.tex
\section{Introduction}
\label{sec:introduction}
% ══════════════════════════════════════════════════════════════════
Jets~\cite{Salam:2010nqg} in High Energy Physics (HEP) are a collection of collimated sprays of stable particles, produced as the relic of hard scattered processes involving partons (viz. splitting and fragmentation of quarks and gluons, followed by hadronization), as well as hadronic decay of other Standard Model (SM) particles like top quark ($t$), heavy flavor quarks ($b,c$), electroweak bosons ($W/Z$) and the Higgs ($H$). 
The task of tracing down the primary truth origin of the particle cascade, which eventually forms a jet, is known as \emph{Jet Tagging}. It is a foundational task in HEP, whose utility spans across most of the physics studies in a hadron collider environment, such as the CERN Large Hadron Collider (LHC), including a wide range
of precision measurements and new-physics
searches~\cite{Kogler:2018hem,LARKOSKI20201}. Deep learning has transformed
this task to new frontiers. Successive architectures, starting from CNNs using jet
images~\cite{deOliveira:2015xxd} to graph neural networks~\cite{Qu:2019gqs,Thais:2022iok,Shlomi:2020gdn} to the
Particle Transformer~\cite{Qu:2022mxj}, have produced classifiers whose
performance substantially exceeds that of any single hand-crafted physics driven observable, such as N-subjettiness \cite{Thaler:2010cxa,Thaler:2011gf}.

The identification of highly boosted top quarks from their hadronic decay
products is one of the central classification tasks in collider physics at
the LHC. A boosted top quark
decaying as $t \to W b \to q\bar{q}b$ produces a single large radius jet
whose constituents encode the kinematics of a three-body decay. The
discrimination of such jets from background jets initiated by light quarks
and gluons is a classification task that has been extensively studied with
both expert designed jet-substructure observables, including
$N$-subjettiness and energy correlation
functions~\cite{Larkoski:2013eya,Moult:2016cvt}. Modern deep learning
architectures, that operate directly on the particle four momenta, 
consistently outperform these expert designed taggers, established extensively on benchmark
datasets~\cite{Kasieczka:2019dbj}.

The Particle Transformer (\ParT{})~\cite{Qu:2022mxj} is among the most widely used jet-tagging
architectures currently available. It augments the standard
Transformer~\cite{2017arXiv170603762V} self-attention mechanism with a learned
pairwise interaction bias derived from pairwise features
of every particle pair, allowing each attention head to express a learned
attention pattern that is informed by the physical structure of the jet.
Despite the strong empirical performance of such models, the internal
computation by which they reach their classification decision has remained
opaque. Explaining these inner workings via analysis of attention maps, Shapley values and information flow is an active area of research in machine learning-based jet studies.

Mechanistic interpretability~\cite{elhage2021mathematical,2022arXiv221100593W,2023arXiv230414997C} is a
research program that aims to reverse-engineer the computations performed
by neural networks by identifying the minimal set of components, called a
\emph{circuit}, that is causally responsible for a given behavior, and by
characterizing the algorithm those components implement. Originally
developed in the context of language models, this program has produced a
toolkit of activation patching, path patching, logit lens, and linear
probing techniques that has not previously been applied systematically to
jet physics classifiers. The present work bridges this gap.

We train a small Particle Transformer (4 particle attention layers, 4
attention heads per layer, embedding dimension $128$, approximately $1.3$M trainable parameters) on a random subset of the Top Quark Tagging reference
dataset~\cite{Kasieczka:2019dbj,kasieczka_2019_2603256} and apply a complete mechanistic
interpretability analysis. The contributions of this work are as follows.

\begin{enumerate}
    \item We identify a six-head circuit that recovers $97.3\%$ of full model
    AUC and is shown to be significantly more performing than randomly
    sampled six-head subsets at the $96$-th percentile of the random baseline
    distribution.

    \item We establish the causal structure of the circuit through path
    patching, distinguishing a single \emph{source} head, three \emph{relay}
    heads, a \emph{secondary source} head, and a single \emph{readout} head.
    The sign pattern of direct effects is consistent across two on-manifold
    corruption strategies, and we document a structural incompatibility
    between off-manifold (Gaussian) corruption and the standard recovery-score
    formulation that we expect to be relevant for any kinematically narrow
    physics dataset.

    \item Linear probing of the residual stream against the family of
    classical jet-substructure observables reveals two distinct
    findings. First, the model preferentially encodes the energy
    correlator basis over the $N$-subjettiness basis, with the
    advantage persisting after residualization of $D_2^{(\beta=1)}$
    and $\tau_{32}$ against jet mass. Second, within the energy
    correlator basis the model preferentially encodes 2-prong
    observables ($C_2^{(\beta=1)}$, $D_2^{(\beta=1)}$) over 3-prong
    observables ($C_3^{(\beta=1)}$, $N_3^{(\beta=1)}$), despite the
    latter being the canonical targets for a top tagger. It appears that the model
    has implicitly factorized the 3-prong top-tagging task into the
    more accessible 2-prong sub-task of identifying the hadronic
    $W$-boson decay.

    \item Resolving the standard logit-lens analysis with per-layer trained
    logistic probes, we find that the apparent single-step commitment to a
    classification decision in the first class attention block is more like a
    \emph{basis rotation}: linearly accessible class information is already
    present at $\mathrm{AUC} \approx 0.97$ in the particle attention layers,
    and the class attention block aligns this signal with the basis of the
    final classification head.
\end{enumerate}

The paper is structured as follows. Section~\ref{sec:related} reviews the existing volume of literature with similar goals.

The relevant background on jet substructure and the Particle Transformer
architecture are given in Appendix~\ref{subsec:part}. Section~\ref{sec:mechinterp} reviews mechanistic interpretability. Section~\ref{sec:setup}
describes the experimental setup. Section~\ref{sec:logitlens} presents the
logit-lens analysis and its basis-rotation interpretation.
Section~\ref{sec:circuit} identifies and characterizes the circuit.
Section~\ref{sec:representations} analyses the physical content of the
residual stream, including the mass-residualized probe comparison.
Section~\ref{sec:features} examines the interaction features driving each
head. Section~\ref{sec:discussion} discusses the results in details and their
implications. Finally we conclude in Section~\ref{sec:conclusion}.

%% file: sections/existing_literature.tex
\section{Existing Literature}
\label{sec:related}

Interpretability and explainability of machine learning models for collider physics have developed into an active subfield, motivated both by the desire to understand what these models compute and by the practical need to estimate systematic uncertainties on their outputs. Existing work falls broadly into three categories: post-hoc attribution methods that assign importance scores to inputs of a trained model, interpretable-by-construction architectures that constrain the model's function class, and recent attention pattern analyses of transformer-based jet taggers. To our knowledge, no prior work in collider physics has applied the full circuit level toolkit of mechanistic interpretability involving zero ablation, path patching, logit lens, and per-layer trained probes operating jointly on a single trained model, i.e. a jet classifier.

\subsection{Post-hoc attribution and explainability methods}

The earliest interpretability efforts for jet taggers were predominantly attribution based, identifying which input features or constituents most strongly influence the model's prediction. Layer wise relevance propagation was applied to deep neural network jet taggers in combination with expert variables~\cite{agarwal2021explainable}, confirming that learned models attribute importance to substructure observables consistent with theoretical expectations. A more detailed study of interpretability for DNN-based top taggers, combining layer wise relevance propagation, saliency maps, and gradient based attribution, was presented in Ref.~\cite{khot2023detailed}; this is the closest predecessor to the present work in topic, but it employs an entirely different methodological toolkit. Shapley value attribution was applied to boosted top tagging in Ref.~\cite{bhattacherjee2024boosted}, yielding per-input contributions to the classification score that satisfy the standard cooperative game axioms of additive feature attribution. Graph specific attribution methods, viz. GNNExplainer, GNNShap, and Grad-CAM, operating on the Lund jet plane were systematically compared recently in Ref.~\cite{Patel:2026zbq}. Beyond top tagging, attribution methods have been applied to particle-flow reconstruction~\cite{mokhtar2021explaining}, to multi-method explainability of muonic-pattern recognition~\cite{maglianella2023convergent} and to uncertainty quantification with evidential deep learning for jet identification, where epistemic uncertainty was used as a proxy for out-of-distribution detection~\cite{khot2025evidential}. The question of whether graph neural networks genuinely learn classical jet-substructure observables was addressed in Ref.~\cite{mokhtar2022graph} by training auxiliary regressors to predict observables from intermediate representations. This is a probing style methodology that is in spirit closest to the linear probing of Section~\ref{sec:representations} of the present work, but which was not coupled to a causal circuit analysis.

What distinguishes attribution methods from the present analysis is that they report \emph{which inputs matter}, but not \emph{how} the network combines them. A high relevance, saliency, or Shapley score for a particular constituent does not localize the computation in network space, does not establish causal information flow between internal components, and does not isolate a sparse sub-network that is itself sufficient for the behavior. The mechanistic interpretability program employed in the present work is designed precisely to address these complementary questions.

\subsection{Interpretable-by-construction architectures}

A second strand of work seeks interpretability not by analyzing a trained black box but by constraining the architecture so that its computation is transparent by design. Binary JUNIPR~\cite{andreassen2019binary} phrases classification as a probabilistic model over the parton shower history, producing a discriminator whose internal log-likelihood ratios admit a direct physical reading at every node of the shower tree. Symbolic regression has been applied to extract closed form expressions for angular coefficients in Drell-Yan production~\cite{bendavid2026angular}, yielding equations that are interpretable by inspection at the cost of restricting the function class the model can express. The IRC-safe equivariant feature extractor of Ref.~\cite{konar2026stable} enforces infrared and collinear safety as an architectural constraint, ensuring that learned representations respect the same physical invariance properties as analytic substructure observables. Interpretable anomaly detection in jet substructure~\cite{bradshaw2022creating} likewise restricts the detector to a small expert designed feature space so that the learned decision boundary can be inspected directly. More recently, the Mixture-of-Experts graph transformer architecture of Ref.~\cite{genovese2025mixture} introduces interpretability through gating. In this method, each expert specializes in a distinct kinematic regime, and the gating distribution serves as an interpretable summary of the learned partitioning of input space.

These approaches buy interpretability through architectural restriction. The present work occupies a complementary position: rather than constraining the architecture, we take a state-of-the-art unrestricted Particle Transformer and ask whether its internal computation can be reverse engineered after training. The two strategies may turn out to be complementary. Insights from circuit level analyses such as the one presented here can in principle inform the design of more interpretable physics aware architectures, and the source $\rightarrow$ relay $\rightarrow$ readout structure identified in Section~\ref{sec:circuit} would be a natural template for an interpretable-by-construction successor.

\subsection{Transformer-specific interpretability for jet tagging}

The most direct neighbors of the present work are recent analyses of transformer-based jet taggers. The physical content of ML quark-vs-gluon discriminators has been characterized in Ref.~\cite{vent2026physics} through correlation studies and feature importance analyses, establishing which observables trained taggers exploit without localizing the computation to specific architectural components. The interaction aware Transformer of Ref.~\cite{Esmail:2025kii} introduces architectural modifications informed by physics priors and inspects the resulting attention patterns, providing input-attribution-style insight without causal interventions on the trained model. Closer still are Refs.~\cite{Wang:2024rup, Legge:2025cnm, Erdmann:2025xpm}: Ref.~\cite{Wang:2024rup} interprets transformers for jet tagging through attention visualization and clustering of head behaviors, identifying head-specialization patterns; the follow-up of Ref.~\cite{Legge:2025cnm} studies the empirical sparsity of attention in the Particle Transformer specifically, characterizing which particle pairs receive non trivial attention weight as a function of layer and head; and Ref.~\cite{Erdmann:2025xpm} applies probing and feature attribution methods to a transformer trained on physics data and asks what physical content the trained model has acquired.

These transformer specific analyses share with the present work a focus on the internal structure of attention based jet classifiers, but they stop short of the causal toolkit deployed here. None performs path patching to establish causal information flow between heads, none distinguishes direct from indirect effects through systematic intervention, and none confronts the basis rotation ambiguity in the standard logit lens that the per layer trained probes of Section~\ref{subsec:basisrotation} resolve. The present work extends these efforts by combining attention pattern analysis, consistent in spirit with Refs.~\cite{Wang:2024rup, Legge:2025cnm}, with the causal intervention methodology of mechanistic interpretability, and by closing the loop with a quantitative comparison between the encoded representation and the classical energy-correlator basis of jet substructure.

%% file: sections/mechanistic_interpretability.tex
\section{Mechanistic interpretability}
\label{sec:mechinterp}

Mechanistic interpretability~\cite{elhage2021mathematical} seeks to decompose a neural
network's computation into human interpretable components. The central
objects of study are \emph{circuits}, a subset of the computational graph
(here attention heads and their connections) that are causally responsible
for a specified behavior. Several intervention tools are used in the
present work.

\paragraph{Zero ablation.} 
In this method, the contribution of a component is removed by setting its output to zero 
either by overwriting its activation or in the case of head scaling
architectures, by setting the head's scaling weight to zero. The resulting
drop in some scalar performance measure quantifies how structurally important the component is.

\paragraph{Path patching.}
The prescription of this method is as following : the model is run on a \emph{clean} input and a \emph{corrupted} input. The
output of a chosen component is recorded on the clean run and substituted
into the corresponding position of the corrupted run, with all other
components processing the corrupted input. Let
$\mathrm{LD} = \log p_{\mathrm{top}} - \log p_{\mathrm{QCD}}$ denote the
logit difference (with \(p_{\mathrm{top} }\) and \(p_{\mathrm{QCD} }\) being the output probability score of the tagger for the binary classification between two classes of \emph{top} and \emph{QCD}), which serves as the scalar measure of the model's
confidence in the top hypothesis. The \emph{direct effect} of a head $(l,h)$
is the recovery score
\begin{equation}
    \mathrm{DE}(l,h)
    \;=\;
    \frac{\mathrm{LD}_{\mathrm{patched}}(l,h) - \mathrm{LD}_{\mathrm{corrupt}}}
         {\mathrm{LD}_{\mathrm{clean}} - \mathrm{LD}_{\mathrm{corrupt}}}.
    \label{eq:directeffect}
\end{equation}
where $\mathrm{LD}_{\mathrm{patched}}(l,h)$ denotes the logit difference
of the model run on the corrupted input with the activation of head
$(l,h)$ replaced by its value from the clean run, and
$\mathrm{LD}_{\mathrm{clean}}$, $\mathrm{LD}_{\mathrm{corrupt}}$ denote
the logit differences on the clean and corrupted runs respectively
without any patching.
A direct effect of $+1$ corresponds to the patched run reproducing the
clean logit difference exactly; values larger than $+1$ correspond to
super-recovery, in which the patched activation drives the logit
difference further past its clean-run value than the clean run itself.
On the dataset studied here the within-batch replacement corruption
yields $\overline{\mathrm{LD}}_{\mathrm{corrupt}} >
\overline{\mathrm{LD}}_{\mathrm{clean}}$ (corrupted top jets are mildly
more top-like, since the replacing particles are themselves drawn from
a $50\%$-top sample; see Section~\ref{subsec:pathpatching}), so the
denominator of Eq.~(\ref{eq:directeffect}) is negative. Within this
sign convention, a negative direct effect indicates that transplanting
the head's clean-run activation drives the patched run's logit
difference \emph{above} the unpatched corrupted baseline, i.e., that
the head's clean-run output is incompatible with the rest of the
corrupted run.

The \emph{path effect} between a source head $(l_s, h_s)$ and a target head
$(l_t, h_t)$ with $l_s < l_t$ is the $L^2$ norm of the difference between
the target head's output under (i) the corrupted run and (ii) the corrupted
run with only the source head's output replaced by its clean-run value.
This isolates the contribution that flows from the source head to the
target head along the residual stream and through the intervening
architecture.

\paragraph{Logit lens.}
The logit lens~\cite{nostalgebraist2020logitlens} projects an intermediate
representation through the final layer normalization and the trained
classification head, producing a pseudo prediction at each layer. The area under the curve (AUC)
of the resulting pseudo prediction quantifies the amount of class
information that is linearly accessible to the trained head at that depth.
A limitation of the logit lens is that it assumes the intermediate
representation is expressed in the same basis as that on which the head
was trained; we revisit this point in Section~\ref{sec:logitlens}.

\paragraph{Linear probing.}
Linear probing~\cite{2016arXiv161001644A} method fits a small linear model (Ridge regression~\cite{2015arXiv150909169V}
for continuous targets, logistic regression for binary targets) to predict
a target observable from the intermediate representation. The held out
coefficient of determination $R^2$ (or AUC, for binary targets) measures the
linear accessibility of that observable in the representation. As with the
logit lens, the result is a lower bound on the true information content,
because non linearly encoded information is not detected.

\begin{figure}[t]
    \centering
    \includegraphics[width=0.95\textwidth]{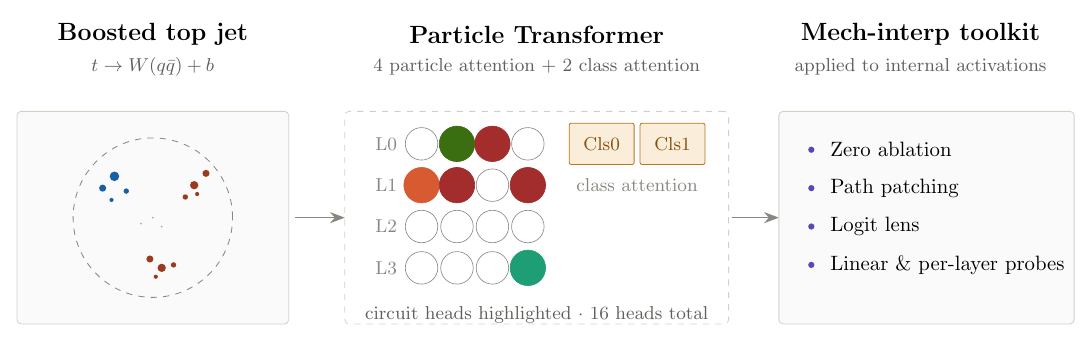}
    \caption{Schematic overview of the analysis pipeline. A boosted top jet
($t \to W(q\bar{q}) + b$, left) is processed by a Particle Transformer
with four particle attention layers and two class attention layers, for
a total of sixteen attention heads (center); the six heads identified
in this work as constituting the classification circuit are filled,
with the remaining ten heads shown as empty circles. The mechanistic
interpretability toolkit applied to the model's internal activations
(right) comprises zero ablation, path patching, the logit lens, and
linear and per-layer probes of the residual stream. The detailed
direct-effect values and information-flow paths within the circuit
are presented in Figure~\ref{fig:circuit}.}
    \label{fig:mech_interp_fig}
\end{figure}

The mechanistic interpretability of the \ParT{} is shown in the Figure~\ref{fig:mech_interp_fig}.

%% file: sections/exp_setup.tex
% ─────────────────────────────────────────────────────────────────────────────
\section{Experimental Setup}
\label{sec:setup}
% ─────────────────────────────────────────────────────────────────────────────

\subsection{Dataset}
\label{subsec:dataset}

We use the Top Quark Tagging reference dataset of
Refs.~\cite{Kasieczka:2019dbj,kasieczka_2019_2603256}, which consists of simulated jets from
$pp$ collisions at $\sqrt{s} = 14$~TeV. Signal jets are initiated by
hadronic top decays, $t \to W b \to q\bar{q}b$; background jets are
initiated by light quarks and gluons. Events are simulated with PYTHIA~8~\cite{Bierlich:2022pfr} and
detector effects are modeled by a fast parametric simulation. Both classes
are selected with jet transverse momentum $p_T \in [550, 650]$~GeV and
pseudo-rapidity $|\eta| < 2$. Jets are reconstructed with the anti-$k_T$
algorithm~\cite{Cacciari:2008gp} using FastJet~\cite{Cacciari:2011ma} with radius
parameter $R = 0.8$. Each jet is represented by up to $200$ constituent
particles ordered by decreasing $p_T$, with zero-padding for jets containing
fewer constituents.

We use $200{,}000$ jets for training, $50{,}000$ for validation, and
$50{,}000$ for testing, with a $50\%$ signal fraction in all splits. All
mechanistic interpretability analyses use the test set unless otherwise
stated. The full dataset is not used primarily due to computationally intensive nature of the study.

\subsection{Model architecture}
\label{subsec:model}

We train a \ParT{} with $4$ particle attention layers, $2$
class attention layers, $4$ attention heads per layer, embedding dimension
$d = 128$, feed-forward expansion ratio $4$, and pairwise interaction
embedding dimensions $[64, 64, 64, 4]$ (the final dimension matching the
number of attention heads). The total number of trainable parameters is
approximately $1.3$M. The necessary details of the \ParT{} architecture can be found in Appendix~\ref{subsec:part}.

The per-particle input features for each constituent are: relative
pseudo-rapidity $\Delta\eta$, relative azimuthal angle $\Delta\phi$,
$\log p_T$, $\log E$, $\log(p_T / p_{T,\mathrm{jet}})$,
$\log(E / E_{\mathrm{jet}})$, and $\Delta R$ from the jet axis. The pairwise
interaction MLP receives the four pairwise features in
Eqn.~\ref{eqn:pairfeat} computed from the constituent four-momenta.

We use two related indexing conventions throughout this paper.
Attention heads are labelled $\mathrm{L}\ell\mathrm{H}h$, where 
$\ell \in \{0, 1, 2, 3\}$ indexes the four particle attention layers and 
$h \in \{0, 1, 2, 3\}$ indexes the four heads within each layer.
Residual stream depths are labelled $\mathrm{L}d$, where 
$d \in \{0, 1, 2, 3, 4\}$: $\mathrm{L}0$ denotes the initial particle 
embedding prior to any attention layer, and $\mathrm{L}d$ for $d \geq 1$ 
denotes the residual stream at the output of the $d$-th particle 
attention layer. The output of the layer containing heads 
$\mathrm{L}\ell\mathrm{H}h$ therefore appears in the residual stream at 
depth $\mathrm{L}(\ell+1)$; for example, the primary source head 
$\mathrm{L}0\mathrm{H}1$ resides in the first particle attention layer, 
and its output is read out from the residual stream at depth $\mathrm{L}1$.

\subsection{Training}
\label{subsec:training}

The model is trained with the AdamW optimizer~\cite{2017arXiv171105101L} with
learning rate $10^{-3}$, $\beta_1 = 0.95$, $\beta_2 = 0.999$,
$\epsilon = 10^{-5}$, and weight decay set to zero. A linear schedule
decays the learning rate from $10^{-3}$ to $10^{-5}$ over the final $30\%$
of training steps after a flat warmup. Training proceeds for $30$ epochs
with batch size $256$, mixed-precision (FP16) computation, and gradient
clipping at $L^2$ norm $1.0$. The best checkpoint is selected by
validation AUC.

We train five independent models with random seeds $0$ through $4$ to assess
reproducibility. The test AUCs across the five seeds are
$\{0.9796, 0.9800, 0.9796, 0.9785, 0.9793\}$, giving a mean and standard
deviation of $0.9794 \pm 0.0005$. This is below the AUC of the original
full scale \ParT{} on the same dataset subset
($\approx 0.985$), as expected for the much smaller model used here, but
already in the regime where transformer-class architectures outperform
classical jet substructure observables. All mechanistic interpretability
analyses use the seed-$0$ model unless otherwise stated; multi seed 
results are reported where relevant.

\subsection{Intervention methods}
\label{subsec:interventions}
In this subsection we discuss in details the several intervention methods that have been used in the present study. 

\paragraph{Zero ablation.}
The Particle Transformer used here implements NormFormer-style head
scaling~\cite{2021arXiv211009456S}, where each attention head's output is multiplied by
a learned scalar $c_{l,h}$ before being summed into the residual stream.
We ablate head $(l,h)$ by setting $c_{l,h} = 0$ during inference, which
removes the head's contribution while leaving all other parameters
unchanged. The importance is measured as the drop in the mean logit
difference over top jets,
\begin{equation}
    I_{\mathrm{zero}}(l,h)
    \;=\;
    \overline{\mathrm{LD}}_{\mathrm{baseline}}
    -
    \overline{\mathrm{LD}}_{\mathrm{ablated}}(l,h),
    \label{eq:Izero}
\end{equation}
averaged over $5{,}000$ top jets and across all five training seeds.

\paragraph{Path patching: corruption strategy.}
Path patching requires a corrupted input that destroys the discriminating
substructure while remaining within the kinematic regime on which the model
was trained. The primary strategy used in this work is \emph{within batch
particle replacement}: each top jet's particles are replaced by those of
another jet in the same batch (offset by half the batch size), preserving
the per-particle feature distribution but breaking the jet level kinematic
correlations. Direct effects are computed over $2{,}000$ test jets; we
verify that the denominator $\mathrm{LD}_{\mathrm{clean}} -
\mathrm{LD}_{\mathrm{corrupt}}$ is non-zero and consistent in sign across
the test sample.

To assess the robustness of the direct effect sign pattern, we additionally
run path patching with a \emph{whole-jet permutation} corruption that
permutes intact jets across the batch (without mixing constituents). This
is a milder corruption that preserves intra-jet correlations but breaks the
top-vs-QCD label assignment.

A third strategy, \emph{Gaussian noise corruption}, was attempted in two
implementations, both of which proved to be structurally incompatible with
the standard recovery-score formulation in
Eq.~(\ref{eq:directeffect}); the underlying reason and its consequences for
physics domain interpretability are discussed in
Section~\ref{subsec:corruption_robustness}.

\paragraph{Circuit masking.}
Finally in this method, for the partial minimality test and the random-baseline analysis, all heads
\emph{outside} a specified circuit are zero-ablated and the test AUC is
evaluated on the full $50{,}000$-jet test set. Bootstrap $95\%$ confidence
intervals on circuit AUCs are computed with $1{,}000$ resamples.

%% file: sections/logitlens.tex
% ─────────────────────────────────────────────────────────────────────────────
\section{When Does the Model Commit to Its Prediction?}
\label{sec:logitlens}
% ─────────────────────────────────────────────────────────────────────────────

\subsection{Logit lens trajectory}
\label{subsec:logitlens_results}
\begin{figure}[h]
    \centering
    \includegraphics[width=0.65\textwidth]{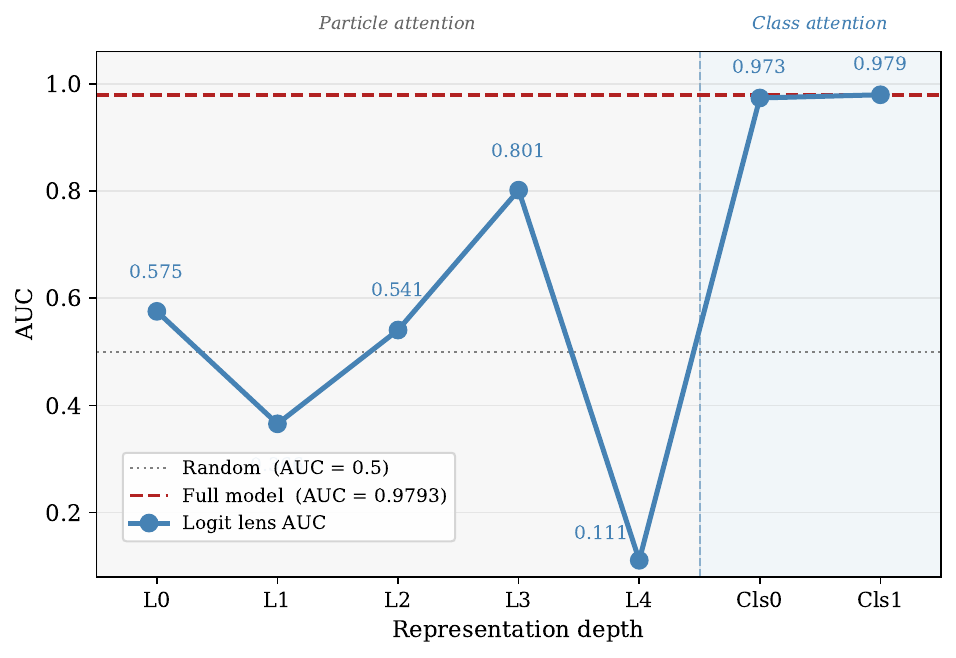}
    \caption{Logit-lens AUC as a function of representation depth. At each
    depth, the mean-pooled particle residual stream is projected through
    the trained final layer normalization and classification head to
    produce a pseudo-prediction. The dashed red line shows the full
    model test AUC of $0.9793$. The shaded region marks the class
    attention blocks. The trajectory through the particle attention
    layers is non-monotone and ends at $\mathrm{AUC} = 0.111$ at L4,
    well below the random baseline; a single-step jump to $0.973$
    occurs at the first class attention block (Cls0).}
    \label{fig:logitlens}
\end{figure}

The logit lens projects the residual stream at each depth through the trained final classification head, producing a pseudo-prediction whose AUC quantifies how much class-discriminating information is linearly accessible to the trained head at that layer.
Figure~\ref{fig:logitlens} shows the logit-lens AUC at each depth in the
network: the initial particle embedding (L0), the four particle attention
layers (L1--L4), and the two class attention blocks (Cls0, Cls1).

The trajectory through the particle attention layers is highly non-monotone:
$0.575 \to 0.366 \to 0.541 \to 0.801 \to 0.111$. The AUC at L4, the output
of the last particle attention layer, is $0.111$, substantially below the
random baseline of $0.5$. This is followed by a jump of $+0.862$ at the
first class attention block (Cls0), with the second class attention block
(Cls1) adding only $+0.006$.

\begin{figure}[H]
    \centering
    \includegraphics[width=0.65\textwidth]{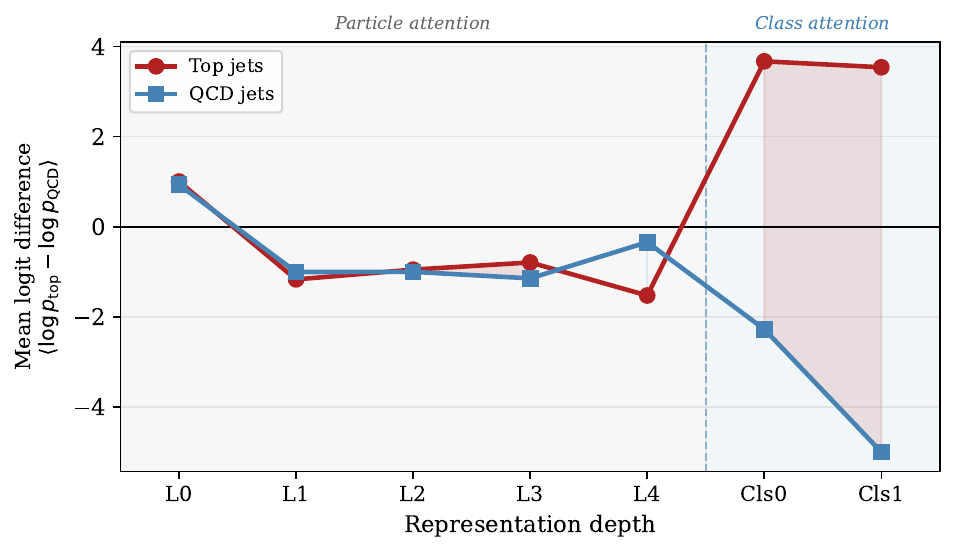}
    \caption{Mean logit difference $\langle \log p_{\mathrm{top}} -
    \log p_{\mathrm{QCD}} \rangle$ as a function of representation depth,
    evaluated separately for top jets (red) and QCD jets (blue) over
    $10{,}000$ test jets. Through all particle attention layers the two
    classes have similar and near-zero logit differences. A sharp class
    separation emerges only at the first class attention block (Cls0).}
    \label{fig:logitlens_ld}
\end{figure}

Figure~\ref{fig:logitlens_ld} shows the corresponding mean logit difference
evaluated separately for top and QCD jets. The two classes have similar
and near-zero logit differences throughout the particle attention layers
and diverge sharply at Cls0, with top jets reaching a mean logit difference
of approximately $+3.7$ and QCD jets dropping to approximately $-2.5$.

Read at face value, these observations suggest that no class-discriminating
signal is present in the particle attention layers and that the entire
classification decision is computed within Cls0. This conclusion would,
however, be premature: the logit lens projects intermediate representations
through a layer normalization and classification head that were trained
specifically on the output of the class attention stack, and a low
logit-lens AUC at an intermediate layer can equally well indicate that the
information is present but in a basis incompatible with the trained head.
We resolve this ambiguity in the next subsection.

\subsection{The class attention block as a basis rotation}
\label{subsec:basisrotation}

To distinguish absent class information from misaligned class information,
we train a per-layer logistic regression probe directly on each layer's
mean-pooled representation, with the same input-output pairs used for the
logit lens. The probe is a single linear layer with its own implicit
normalization (a standard scaler) and is fitted to maximize binary
classification accuracy on a $7{,}000$-jet training subset of the test set,
with AUC evaluated on the remaining $3{,}000$ jets.

\begin{figure}[t]
    \centering
    \includegraphics[width=0.7\textwidth]{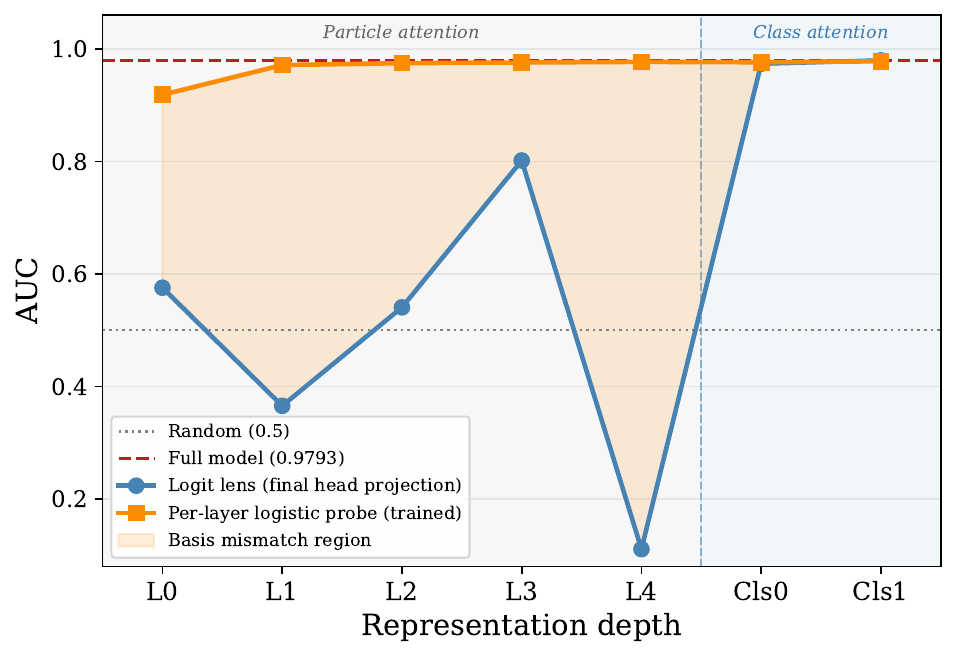}
    \caption{Comparison of logit-lens AUC (blue circles) and per-layer
    trained logistic probe AUC (orange squares) as a function of
    representation depth. The shaded region between the two curves
    quantifies the basis mismatch between the layer's representation
    and the trained classification head. Class-discriminating
    information is essentially complete by L1 ($\mathrm{AUC} = 0.971$)
    and remains so through L4 ($\mathrm{AUC} = 0.977$), even where the
    logit lens reports values below random ($\mathrm{AUC} = 0.111$ at
    L4). The two metrics agree at the class attention blocks
    (Cls0, Cls1).}
    \label{fig:basisrotation}
\end{figure}

Table~\ref{tab:basisrotation} compares the logit-lens AUC and the per-layer
probe AUC at each depth, and Figure~\ref{fig:basisrotation} shows the
comparison graphically.

\begin{table}[h]
    \centering
    \caption{Logit-lens AUC and per-layer trained logistic probe AUC at
    each representation depth, evaluated on the same $10{,}000$ test
    jets. The logit lens uses the trained final classification head;
    the per-layer probe is trained from scratch on each layer. A large
    discrepancy at a particle attention layer indicates that
    class-discriminating information is present in the representation
    but in a basis to which the trained head is not aligned.}
    \label{tab:basisrotation}
    \begin{tabular}{lccc}
        \toprule
        Depth & Logit lens & Per-layer probe & Difference \\
        \midrule
        L0   & $0.575$ & $0.918$ & $+0.343$ \\
        L1   & $0.366$ & $0.971$ & $+0.605$ \\
        L2   & $0.541$ & $0.974$ & $+0.434$ \\
        L3   & $0.801$ & $0.976$ & $+0.174$ \\
        L4   & $0.111$ & $0.977$ & $+0.866$ \\
        Cls0 & $0.973$ & $0.976$ & $+0.003$ \\
        Cls1 & $0.979$ & $0.978$ & $-0.002$ \\
        \bottomrule
    \end{tabular}
\end{table}

The per-layer probe AUC reaches $0.971$ already at L1 and remains in the
range $0.971$-$0.977$ throughout the particle attention layers. At L4,
where the logit lens reports $\mathrm{AUC} = 0.111$, the trained per-layer
probe achieves $\mathrm{AUC} = 0.977$, only marginally lower than the full
model's $0.9793$. The two metrics agree to within $0.003$ at Cls0 and
Cls1.

The mechanistic interpretation that emerges is qualitatively different
from the standard logit-lens reading. The class-discriminating signal
saturates within the first particle attention layer; the residual stream
of L1-L4 contains essentially all the linearly accessible class
information that the network ever produces. The role of the first class
attention block is then a \emph{basis rotation}: a single global attention
operation that reorients the latent class signal from the particle
attention basis (orthogonal to, and partially anti-aligned with, the
trained head) into the basis on which the head was trained. The dramatic
$+0.862$ logit-lens jump at Cls0 measures the magnitude of this rotation,
not a discontinuous gain in information content.

This refinement is consistent with the linear probe analysis of
Section~\ref{sec:representations}, which finds that the binary $\mathrm{is\_top}$
target is encoded with similar $R^2$ at L4 and at the class attention
blocks. It also clarifies the role of the second class attention block
Cls1, which adds essentially no logit-lens AUC ($+0.006$) over Cls0:
once the basis has been aligned, no further computation is necessary.

%% file: sections/circuit_identification.tex
% ─────────────────────────────────────────────────────────────────────────────
\section{Circuit Identification}
\label{sec:circuit}
% ─────────────────────────────────────────────────────────────────────────────
Having established that the class discriminating signal is constructed within the particle attention stack rather than the class attention block, we now ask which of the model's sixteen attention heads are responsible for this computation and how information flows between them. We address this in three stages: zero ablation to identify a candidate set of important heads, path patching to establish the causal structure connecting them, and minimality and random-baseline tests to confirm that the resulting circuit is both sufficient and non trivial.

\subsection{Head importance from zero ablation}
\label{subsec:importance}
\begin{figure}[t]
    \centering
    \includegraphics[width=0.6\textwidth]{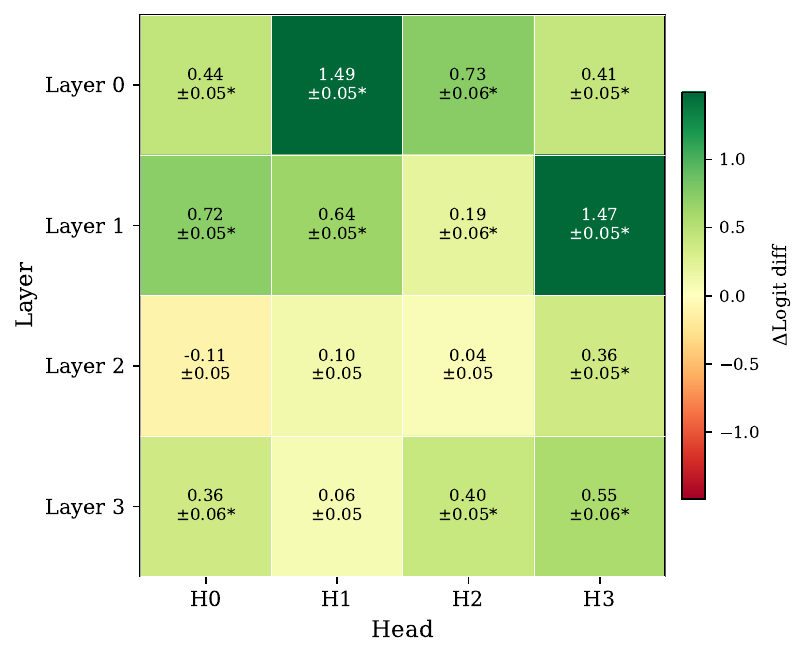}
    \caption{Head importance measured by zero ablation, averaged over
    five independent training seeds. Each cell shows the mean and
    standard deviation of $I_{\mathrm{zero}}(l,h)$, with asterisks
    marking heads significant at the $2\sigma$ level. The model AUC
    is consistent across seeds ($0.9794 \pm 0.0005$). The two most
    important heads are L0H1 ($1.49 \pm 0.05$) and L1H3 ($1.47 \pm
    0.05$).}
    \label{fig:headimportance}
\end{figure}

We begin by ranking the sixteen attention heads by their structural importance, measured as the drop in mean logit difference when each head's output is set to zero.
Figure~\ref{fig:headimportance} reports the zero-ablation importance
$I_{\mathrm{zero}}(l,h)$ for all $16$ attention heads, averaged over the
five training seeds.

Two heads stand out clearly: L0H1 with $I_{\mathrm{zero}} = 1.491 \pm
0.051$ and L1H3 with $I_{\mathrm{zero}} = 1.469 \pm 0.054$. The standard
deviations across seeds lie in the range $0.050$--$0.056$, and the
ranking of the heads is consistent across all five seeds. Twelve of the
sixteen heads are significant at the $2\sigma$ level; the four
non-significant heads are L2H0 ($-0.113 \pm 0.054$), L2H1, L2H2, and L3H1,
all in Layer~2 or in a single Layer~3 head. The top six heads by
importance are L0H1, L1H3, L0H2, L1H0, L1H1, and L3H3. These constitute
the candidate circuit studied in the remainder of this section.

The cross-seed stability of the importance ranking is a non-trivial
feature: a circuit identified in a single trained model could in principle
be an artifact of that particular random initialization. Here, the
identity of the top-six heads varies only in ordering across seeds and
the rankings are stable to within their statistical uncertainties. This
provides a first piece of evidence that the circuit reflects a property
of the training objective and dataset rather than the random
initialization.

\subsection{Causal structure from path patching}
\label{subsec:pathpatching}

Zero ablation measures the structural importance of each head but does
not distinguish between heads that carry discriminating information
independently and heads that are important only in the context of other
heads. Path patching provides this distinction by causally intervening
on individual head outputs.

We perform path patching on the seed-$0$ model with the within-batch
replacement corruption, computing direct effects over $2{,}000$ test
jets. The clean and corrupted mean logit differences are
$\overline{\mathrm{LD}}_{\mathrm{clean}} = +0.021$ and
$\overline{\mathrm{LD}}_{\mathrm{corrupt}} = +0.089$, giving the
denominator $\overline{\mathrm{LD}}_{\mathrm{clean}} -
\overline{\mathrm{LD}}_{\mathrm{corrupt}} = -0.068$. The negative sign
indicates that on the corrupted runs the model becomes mildly more
confident in the top hypothesis, a kinematic consequence of replacing a
top jet's particles with those of another jet that is itself drawn from
a $50\%$-top dataset.

\begin{figure}[h]
    \centering
    \includegraphics[width=0.55\textwidth]{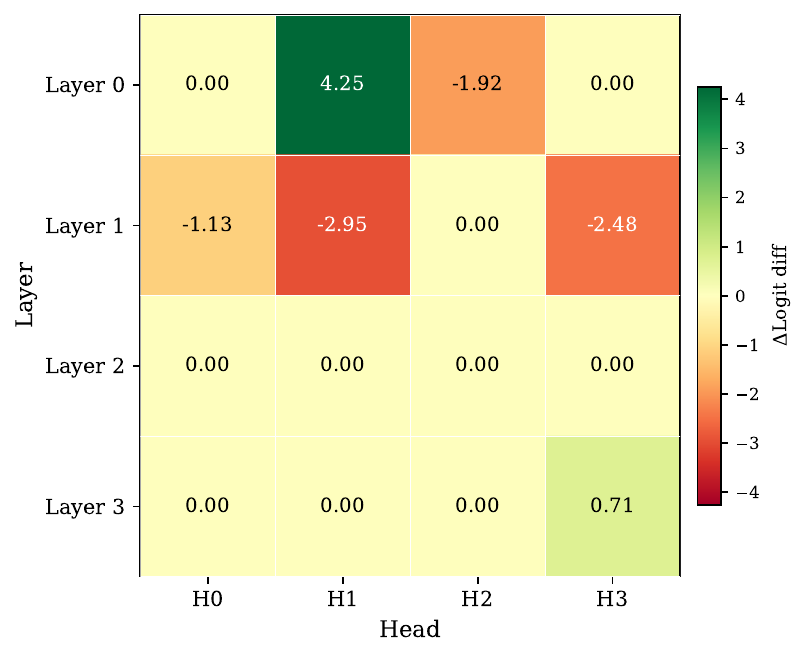}
    \caption{Direct effect (path-patching recovery score) for all $16$
    heads. Heads not in the candidate circuit have direct effect set
    to zero by convention since they were not patched. The dominance
    of L0H1 ($+4.25$), the negative effects of the four Layer~$0$ and
    Layer~$1$ relay-type heads, and the small positive direct effect
    of L3H3 ($+0.71$) are visible.}
    \label{fig:patchimportance}
\end{figure}

Table~\ref{tab:directeffects} lists the direct effect of each circuit
head, and Figure~\ref{fig:patchimportance} shows the full $4\times 4$
direct-effect matrix.

\begin{table}[h]
    \centering
    \caption{Direct effects of the six circuit heads from path patching,
    averaged over $2{,}000$ test jets. Positive values indicate that
    transplanting the head's clean run output into a corrupted run
    recovers the clean logit difference; negative values indicate that
    the transplanted activation is incompatible with the rest of the
    corrupted run. The role assignment is read off the sign and
    magnitude pattern in combination with the path effect graph
    (Section~\ref{subsec:patheffects}) and the representational
    similarity analysis (Section~\ref{subsec:similarity}).}
    \label{tab:directeffects}
    \begin{tabular}{lcl}
        \toprule
        Head & Direct effect & Role \\
        \midrule
        L0H1 & $+4.25$ & Primary source \\
        L0H2 & $-1.92$ & Secondary source \\
        L1H0 & $-1.13$ & Relay \\
        L1H1 & $-2.95$ & Relay \\
        L1H3 & $-2.48$ & Relay \\
        L3H3 & $+0.71$ & Readout \\
        \bottomrule
    \end{tabular}
\end{table}

The direct effect pattern reveals a three-tier structure that we
classify as follows. L0H1 is a \emph{primary source}: a single head whose
clean-run output, transplanted alone, is sufficient to recover most of
the discriminating signal (DE $= +4.25$, i.e., super-recovery).
L3H3 is a \emph{readout} head: its modest positive direct effect
($+0.71$) is consistent with an aggregation operation that benefits from,
but does not produce, the underlying class signal. L0H2 (DE $= -1.92$)
and the three Layer~$1$ heads L1H0, L1H1, L1H3 (DE $= -1.13, -2.95,
-2.48$) all have negative direct effects, indicating that their
clean-run activations are conditional on the upstream signal from L0H1
and that transplanting any one of them in isolation introduces a
representational inconsistency that degrades performance. The cosine
similarity analysis of Section~\ref{subsec:similarity} groups L0H2 with
L0H1 (cosine similarity $0.92$), placing L0H2 in a \emph{secondary
source} role rather than a relay role; this taxonomy is used
consistently throughout the remainder of the paper.

\subsection{Path effects with bootstrap confidence intervals}
\label{subsec:patheffects}

Direct effects identify which heads carry causal weight but do not
specify how information flows between them. This is given by path effects, defined as
the $L^2$ representational delta between corrupted and source-patched
target outputs.

Table~\ref{tab:patheffects} reports the eleven forward path effects in
the candidate circuit, with $95\%$ bootstrap confidence intervals
computed from $500$ bootstrap resamples over $2{,}000$ test jets.
% Figure~\ref{fig:patheffects} displays the same data graphically.

\begin{table}[t]
    \centering
    \caption{Forward path effects (representational deltas) within the
    candidate circuit, with $95\%$ bootstrap confidence intervals over
    $500$ bootstrap resamples on $2{,}000$ test jets. The three
    natural groupings (primary source $\to$ Layer~$1$, secondary
    source $\to$ Layer~$1$, Layer~$1$ $\to$ readout) are pairwise
    non-overlapping, establishing a quantitative ordering of the
    information flow.}
    \label{tab:patheffects}
    \begin{tabular}{lccc}
        \toprule
        Path & Mean $\Delta$ & $95\%$ CI lower & $95\%$ CI upper \\
        \midrule
        L0H1 $\to$ L1H1 & $1.454$ & $1.417$ & $1.494$ \\
        L0H1 $\to$ L1H3 & $1.404$ & $1.360$ & $1.443$ \\
        L0H1 $\to$ L3H3 & $1.351$ & $1.318$ & $1.383$ \\
        L0H1 $\to$ L1H0 & $1.189$ & $1.160$ & $1.217$ \\
        \midrule
        L0H2 $\to$ L1H0 & $0.900$ & $0.881$ & $0.919$ \\
        L0H2 $\to$ L1H1 & $0.858$ & $0.839$ & $0.876$ \\
        L0H2 $\to$ L3H3 & $0.855$ & $0.838$ & $0.872$ \\
        L0H2 $\to$ L1H3 & $0.850$ & $0.833$ & $0.867$ \\
        \midrule
        L1H3 $\to$ L3H3 & $0.754$ & $0.735$ & $0.771$ \\
        L1H0 $\to$ L3H3 & $0.752$ & $0.736$ & $0.769$ \\
        L1H1 $\to$ L3H3 & $0.683$ & $0.668$ & $0.698$ \\
        \bottomrule
    \end{tabular}
\end{table}

The eleven path effects fall into three non-overlapping groups. L0H1
sends path effects of $1.19$--$1.45$ to its four downstream targets
(the three Layer~$1$ relay heads and the L3H3 readout). L0H2 sends
weaker path effects of $0.85$--$0.90$ to the same targets. The Layer~$1$
relay heads in turn send path effects of $0.68$--$0.75$ to the L3H3
readout. The hierarchy primary-source $\succ$ secondary-source $\succ$
relay-to-readout is statistically resolved: the smallest L0H1-sourced
path effect ($1.19$) is more than $25$ standard deviations above the
largest L0H2-sourced one ($0.90$).

\begin{figure}[h]
    \centering
    \includegraphics[width=0.85\textwidth]{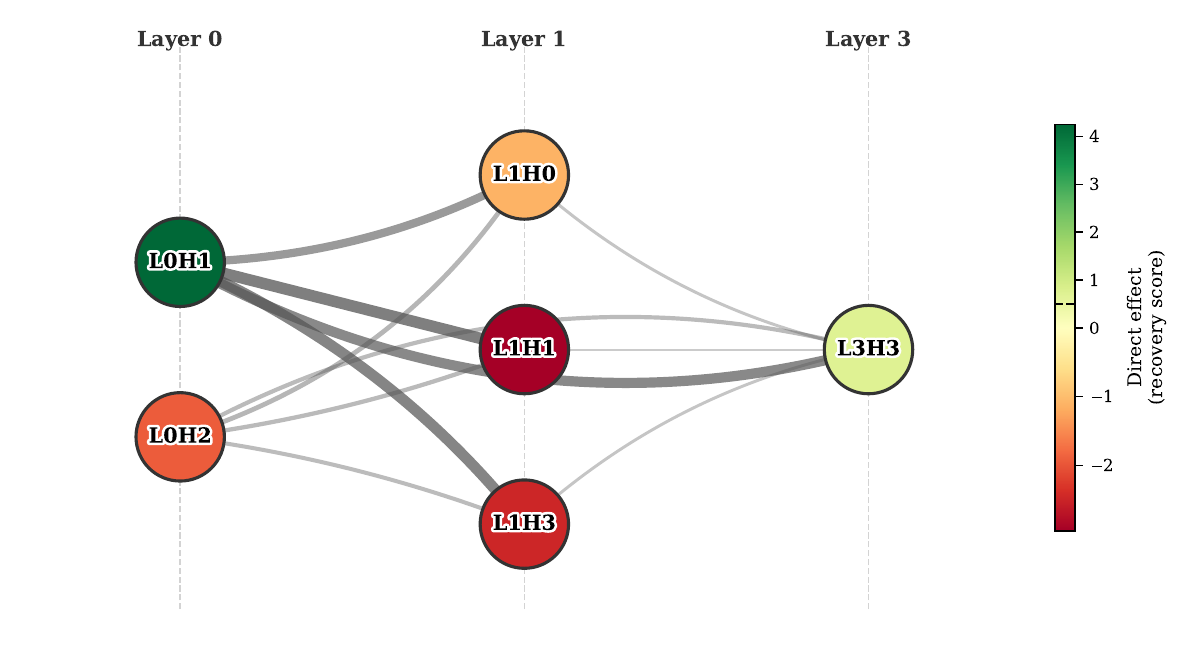}
    \caption{The identified circuit. Nodes represent the six circuit
    heads; node color encodes the direct effect on a diverging scale
    (green: positive; red: negative); edge thickness encodes the
    path-effect magnitude. The circuit has a source-relay-readout
    structure: L0H1 (primary source, $\mathrm{DE} = +4.25$) sends the
    strongest path effects to three Layer~$1$ relay heads, which
    collectively forward information to L3H3 (readout, $\mathrm{DE} =
    +0.71$). L0H2 (secondary source, $\mathrm{DE} = -1.92$) sends
    weaker but still substantial path effects to the same relay heads.
    No Layer~$2$ head appears in the circuit.}
    \label{fig:circuit}
\end{figure}

The structure suggested by these path effects is summarised in
Figure~\ref{fig:circuit}.

\subsection{Minimality and sufficiency}
\label{subsec:minimality}

A true test of minimality for the circuit would require an analysis using methods like ACDC. We do not do such a minimality test in this work. We test the sufficiency of the six-head circuit by
adding heads one at a time in order of decreasing recovery score and
evaluating the masked-circuit AUC on the full test set, with bootstrap
$95\%$ confidence intervals.
Table~\ref{tab:minimality} reports the results.

\begin{table}[t]
    \centering
    \caption{Partial minimality and sufficiency test. Heads are added to the
    circuit in order of decreasing recovery score. At each step, all
    heads outside the current circuit are masked and the test AUC is
    evaluated on the full $50{,}000$-jet test set. Bootstrap $95\%$
    confidence intervals are computed with $1{,}000$ resamples. The
    six-head circuit recovers $97.3\%$ of the full model AUC of
    $0.9793$.}
    \label{tab:minimality}
    \begin{tabular}{clcccc}
        \toprule
        Step & Head & DE & AUC & $95\%$ CI & \% of full model \\
        \midrule
        1 & L0H1 & $+4.25$ & $0.8676$ & $[0.8641, 0.8710]$ & $88.6\%$ \\
        2 & L3H3 & $+0.71$ & $0.8921$ & $[0.8888, 0.8953]$ & $91.1\%$ \\
        3 & L1H0 & $-1.13$ & $0.9054$ & $[0.9026, 0.9079]$ & $92.5\%$ \\
        4 & L0H2 & $-1.92$ & $0.9089$ & $[0.9062, 0.9114]$ & $92.8\%$ \\
        5 & L1H3 & $-2.48$ & $0.9469$ & $[0.9450, 0.9488]$ & $96.7\%$ \\
        6 & L1H1 & $-2.95$ & $0.9527$ & $[0.9510, 0.9544]$ & $97.3\%$ \\
        \bottomrule
    \end{tabular}
\end{table}

\begin{figure}[h]
    \centering
    \includegraphics[width=0.65\textwidth]{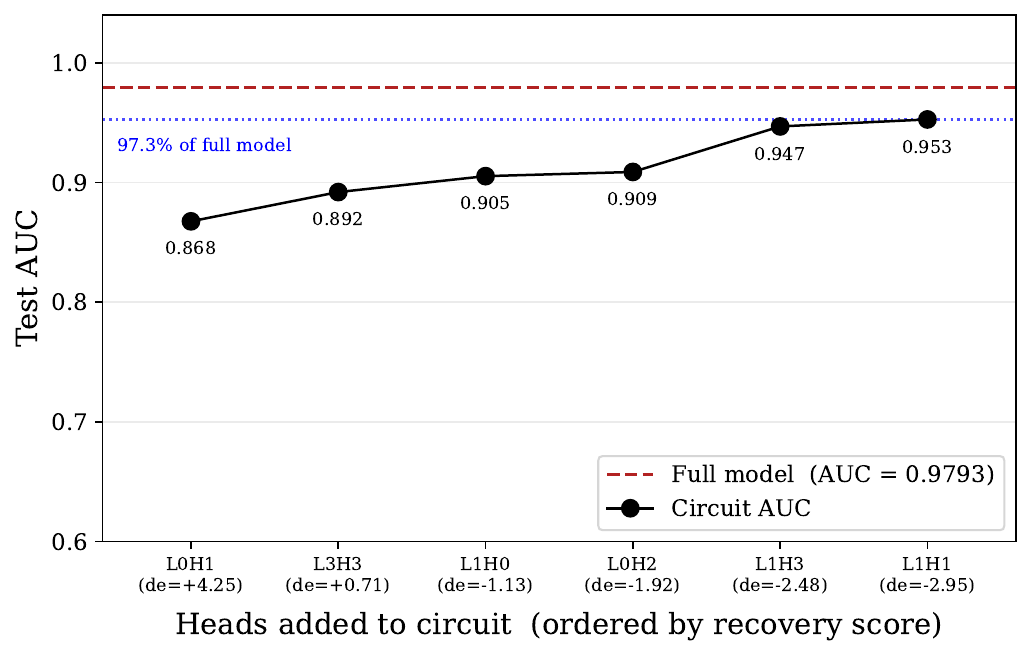}
    \caption{Partial minimality test. Test AUC of the partial circuit as heads
    are added one at a time in order of decreasing recovery score.
    Error bars show the $95\%$ bootstrap confidence interval. The
    dashed red line shows the full model AUC ($0.9793$). The two
    largest single-step gains occur when the primary source L0H1 is
    added (step 1, $+0.868$ AUC) and when the relay head L1H3 is
    added (step 5, $+3.8$ percentage points), the latter only after
    the upstream context from L0H1, L3H3, L1H0, and L0H2 is in place.}
    \label{fig:minimality}
\end{figure}

The trajectory has two notable nonlinearities. L0H1 alone achieves
$88.6\%$ of the full model AUC, consistent with its dominant direct
effect. The most significant non monotonic feature occurs at step~$5$,
when L1H3 is added: the AUC jumps from $0.909$ to $0.947$, an increment
of $+3.8$ percentage points. This is substantially larger than any of
the preceding steps, despite L1H3 having the second-most-negative direct
effect ($-2.48$). The interpretation is that L1H3 functions effectively
only when the upstream context (L0H1, L3H3, L1H0, L0H2) is already in
the circuit; this pattern of conditional sufficiency is the defining
signature of a relay node and distinguishes the present circuit from a
collection of independent components.

\subsection{Random-baseline comparison}
\label{subsec:random}

To assess whether the identified circuit is significantly more performant
than an arbitrary six-head subset, we compare its AUC against a baseline
of $200$ randomly sampled six-head subsets. The remaining heads in each
random subset are masked, and the test AUC is computed on the full test
set with all other settings identical.

\begin{figure}[h]
    \centering
    \includegraphics[width=0.65\textwidth]{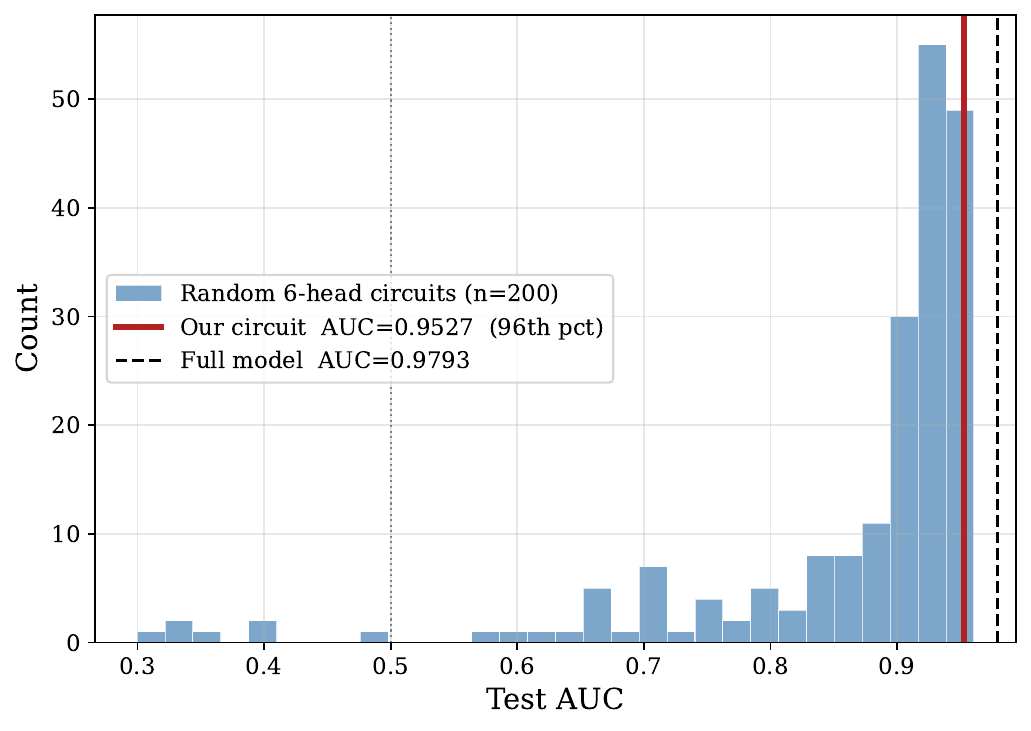}
    \caption{Random-baseline comparison. Test AUC distribution for $200$
    randomly sampled six-head circuits (blue histogram), compared to
    the ablation-identified circuit (red vertical line,
    $\mathrm{AUC} = 0.9527$) and the full model (black dashed line,
    $\mathrm{AUC} = 0.9793$). The identified circuit lies at the
    $96$th percentile of the random distribution. The random
    distribution has substantial spread ($\sigma = 0.125$), with
    several subsets exceeding $\mathrm{AUC} = 0.93$ and others
    falling below $0.65$, indicating significant variation across
    random sparse subsets and a partly redundant learned
    representation.}
    \label{fig:randombaseline}
\end{figure}

The random six-head subsets achieve a mean AUC of $0.869$ with standard
deviation $0.125$, compared to the identified circuit's $0.953$
(Figure~\ref{fig:randombaseline}). The identified circuit lies at the
$96$-th percentile of the random distribution: of the $200$ random
subsets, only $8$ achieve an AUC at or above $0.953$. We refrain from
quoting a parametric $z$-test $p$-value here because the broad spread
of the random distribution makes the empirical percentile a more
directly interpretable summary; what matters is that the identified
circuit is significantly more performant than typical sparse subsets
while not being uniquely so. The substantial spread of the random
distribution, with several random subsets achieving $\mathrm{AUC}$
above $0.93$ but others falling below $0.65$, is informative in its
own right and indicates that the model has learned a somewhat
redundant representation in which many sparse subsets retain
substantial task-relevant information while many others lose it
entirely. The merit of the identified circuit is the combination of
its strong AUC \emph{and} its clean mechanistic interpretation as a
source-relay-readout structure (an interpretation that random subsets
do not, in general, admit).

A complete (though much more computationally intensive) enumeration of all $\binom{16}{6} = 8008$ six-head subsets
would replace the random-baseline percentile with an exact rank.

\subsection{Robustness of the direct-effect sign pattern under alternative
corruption strategies}
\label{subsec:corruption_robustness}

Path-patching results depend on the choice of corruption strategy
(Section~\ref{subsec:interventions}). To test whether the sign pattern
of the direct effects is robust, we re-run the analysis on $2{,}000$
test jets with the whole-jet permutation corruption.
Table~\ref{tab:corruption_robustness} reports the result.

\begin{table}[h]
    \centering
    \caption{Direct effects of the six circuit heads under two
    on-manifold corruption strategies: within batch particle
    replacement (the primary strategy used throughout this work) and
    whole jet permutation. The qualitative sign pattern, viz., positive
    for the primary source L0H1 and the readout L3H3, negative for
    the secondary source L0H2 and all three relay heads, is fully
    consistent across the two strategies. Magnitudes are smaller
    under jet permutation than under replacement, which we attribute
    to the smaller absolute denominator $\mathrm{LD}_{\mathrm{clean}}
    - \mathrm{LD}_{\mathrm{corrupt}}$ achieved by the milder
    corruption.}
    \label{tab:corruption_robustness}
    \begin{tabular}{llcc}
        \toprule
        Head & Role & Replacement & Jet permutation \\
        \midrule
        L0H1 & Primary source   & $+4.25$ & $+3.78$ \\
        L0H2 & Secondary source & $-1.92$ & $-1.72$ \\
        L1H0 & Relay            & $-1.13$ & $-0.67$ \\
        L1H1 & Relay            & $-2.95$ & $-2.51$ \\
        L1H3 & Relay            & $-2.48$ & $-2.17$ \\
        L3H3 & Readout          & $+0.71$ & $+0.66$ \\
        \bottomrule
    \end{tabular}
\end{table}

The sign pattern is fully consistent: L0H1 and L3H3 have positive direct
effects under both strategies; L0H2 and the three relay heads have
negative direct effects under both. Magnitudes are uniformly smaller
under permutation than under replacement, which is consistent with the
smaller absolute value of the denominator $\mathrm{LD}_{\mathrm{clean}}
- \mathrm{LD}_{\mathrm{corrupt}}$ on this milder corruption.

\paragraph{Gaussian noise corruption.}
We additionally explored Gaussian noise corruption in two implementations:
additive noise on the four-momentum components $(p_x, p_y, p_z, E)$, and
additive noise in the kinematic representation $(\log p_T, \eta, \phi)$
followed by reconstruction of physically valid four-vectors satisfying
$E = |\mathbf{p}|$. At every noise scale that we examined (from $0.05$
to $2.0$ times the empirical kinematic standard deviation), Gaussian
corruption drives the model into a strongly QCD-confident regime, with
$\mathrm{LD}_{\mathrm{corrupt}} \ll 0$. This inverts the sign of the
denominator $\mathrm{LD}_{\mathrm{clean}} - \mathrm{LD}_{\mathrm{corrupt}}$,
which is negative under the on-manifold strategies (Section~\ref{subsec:pathpatching})
and would be positive under Gaussian corruption. The recovery scores from
the two regimes are therefore not on a common footing, and Gaussian-corruption
results cannot be combined with the on-manifold ones without redefining
the recovery score. We attribute this to the kinematically narrow training
distribution of the dataset ($p_T \in [550, 650]$~GeV) and to the absence
of data augmentation: off-manifold inputs at any noise scale are sufficiently
far from the training distribution that the model collapses to a fixed,
strongly negative logit difference. We suspect this to be a generic feature
of any physics domain interpretability analysis on a tightly distributed
dataset and we recommend on-manifold corruption strategies as the default
choice.

\subsection{Representational similarity}
\label{subsec:similarity}

The cosine similarity between the mean jet level output representations
of each circuit head pair, computed over $3{,}000$ test jets, is shown in
Table~\ref{tab:cosine}.

\begin{table}[H]
    \centering
    \caption{Cosine similarity between the mean jet-level output
    representations of each circuit head pair, computed over $3{,}000$
    test jets. Two clusters emerge: a Layer~$0$ cluster
    \{L0H1, L0H2\} with cosine similarity $0.92$, and a Layer~$1$
    cluster \{L1H0, L1H1, L1H3\} with pairwise similarities in
    $[0.81, 0.95]$. The L3H3 readout has low similarity to all other
    heads.}
    \label{tab:cosine}
    \begin{tabular}{lcccccc}
        \toprule
        & L0H1 & L0H2 & L1H0 & L1H1 & L1H3 & L3H3 \\
        \midrule
        L0H1 & $1.000$ & $0.922$ & $0.027$ & $0.144$ & $0.069$ & $0.052$ \\
        L0H2 & $0.922$ & $1.000$ & $0.093$ & $0.134$ & $0.112$ & $0.051$ \\
        L1H0 & $0.027$ & $0.093$ & $1.000$ & $0.812$ & $0.952$ & $0.451$ \\
        L1H1 & $0.144$ & $0.134$ & $0.812$ & $1.000$ & $0.901$ & $0.476$ \\
        L1H3 & $0.069$ & $0.112$ & $0.952$ & $0.901$ & $1.000$ & $0.468$ \\
        L3H3 & $0.052$ & $0.051$ & $0.451$ & $0.476$ & $0.468$ & $1.000$ \\
        \bottomrule
    \end{tabular}
\end{table}

Two clusters are immediately visible. The Layer~$0$ heads L0H1 and L0H2
have cosine similarity $0.922$, indicating that their mean output
representations are nearly parallel: this justifies the
\emph{secondary source} taxonomy assigned to L0H2, in contrast to the
qualitatively different (and orthogonal-to-source) representations of
the Layer~$1$ relay heads. The three Layer~$1$ relay heads form a tight
cluster with pairwise cosine similarities $0.812$--$0.952$, consistent
with their playing redundant relay roles. The L3H3 readout has cosine
similarity at most $0.476$ to any other circuit head, consistent with
performing a qualitatively distinct aggregation operation.

The pairwise zero-ablation analysis quantifies the redundancy.
Table~\ref{tab:superadditivity} reports, for all $\binom{6}{2} = 15$
pairs of circuit heads, the joint-ablation damage, the sum of the two
individual-ablation damages, and their difference (the super-additivity).
A negative super-additivity indicates that masking both heads together
causes less damage than the sum of masking each individually, which is
the expected signature of redundancy: when one head is masked, the
other partially compensates by carrying a similar representation.

\begin{table}[t]
    \centering
    \caption{Pairwise zero-ablation analysis for all $\binom{6}{2} = 15$
    pairs of circuit heads. ``Joint'' is the drop in mean logit
    difference when both heads are masked simultaneously; ``Sum'' is
    the sum of the corresponding individual-ablation drops; the
    super-additivity $S$ is their difference. Negative $S$ indicates
    redundancy (joint masking damages less than the sum of individual
    damages); $S \approx 0$ indicates approximately independent
    contributions. The pair \{L0H1, L0H2\} is the most strongly
    redundant ($S = -1.51$), consistent with their high cosine
    similarity (Table~\ref{tab:cosine}); the pairs within the
    Layer~$1$ relay cluster show similarly large redundancy. Pairs
    that combine a Layer~$0$ head with a Layer~$1$ relay head
    (off-cluster pairings) are approximately independent.}
    \label{tab:superadditivity}
    \begin{tabular}{lccc}
        \toprule
        Pair & Joint & Sum & Super-additivity $S$ \\
        \midrule
        \multicolumn{4}{l}{\textit{Within Layer~$0$ (source cluster):}} \\
        \{L0H1, L0H2\} & $0.732$ & $2.244$ & $-1.512$ \\
        \midrule
        \multicolumn{4}{l}{\textit{Within Layer~$1$ (relay cluster):}} \\
        \{L1H0, L1H3\} & $0.738$ & $2.236$ & $-1.498$ \\
        \{L1H1, L1H3\} & $0.638$ & $2.136$ & $-1.498$ \\
        \{L1H0, L1H1\} & $0.638$ & $1.375$ & $-0.738$ \\
        \midrule
        \multicolumn{4}{l}{\textit{L0H1 paired with each remaining circuit head:}} \\
        \{L0H1, L1H3\} & $2.570$ & $3.011$ & $-0.441$ \\
        \{L0H1, L3H3\} & $1.725$ & $2.057$ & $-0.332$ \\
        \{L0H1, L1H1\} & $1.841$ & $2.150$ & $-0.309$ \\
        \{L0H1, L1H0\} & $1.943$ & $2.250$ & $-0.307$ \\
        \midrule
        \multicolumn{4}{l}{\textit{Cross-cluster (Layer~$0$ $\times$ Layer~$1$ off-cluster):}} \\
        \{L0H2, L1H0\} & $1.293$ & $1.469$ & $-0.177$ \\
        \{L0H2, L1H1\} & $1.328$ & $1.369$ & $-0.041$ \\
        \{L1H3, L0H2\} & $2.250$ & $2.230$ & $+0.020$ \\
        \midrule
        \multicolumn{4}{l}{\textit{With readout head L3H3:}} \\
        \{L1H0, L3H3\} & $1.151$ & $1.282$ & $-0.131$ \\
        \{L0H2, L3H3\} & $1.227$ & $1.276$ & $-0.049$ \\
        \{L1H3, L3H3\} & $2.019$ & $2.043$ & $-0.024$ \\
        \{L1H1, L3H3\} & $1.184$ & $1.182$ & $+0.002$ \\
        \bottomrule
    \end{tabular}
\end{table}

The pattern in Table~\ref{tab:superadditivity} confirms and refines the
representational similarity analysis. The two strongest redundancies
($S = -1.51$ for \{L0H1, L0H2\} and $S = -1.50$ for both \{L1H0, L1H3\}
and \{L1H1, L1H3\}) are precisely the pairs identified as
representational near-parallel by the cosine similarity in
Table~\ref{tab:cosine} (cosine similarities $0.92$, $0.95$, and $0.90$
respectively). All five pairings involving L0H1 show negative
super-additivity ($S$ in $[-1.51, -0.31]$), reflecting that L0H1 is so
dominant that the rest of the circuit partially compensates for its
absence in any pairwise masking. By contrast, cross-cluster pairings
that mix a Layer~$0$ source with an off-source-cluster Layer~$1$ relay
or with the L3H3 readout are close to independent ($|S| \le 0.18$),
indicating that the Layer~$0$ and Layer~$1$ clusters carry largely
non-overlapping computational roles, as already suggested by the
representational orthogonality in Table~\ref{tab:cosine} (cosine
similarities below $0.15$ between the two clusters). The readout head
L3H3, in turn, is approximately independent of every relay and
secondary-source head it is paired with ($S$ in $[-0.13, +0.00]$),
consistent with its qualitatively distinct aggregation role.

\subsection{Robustness across kinematic regimes}
\label{subsec:regimes}

Table~\ref{tab:regimes} compares the full model and circuit-only AUC in
bins of jet multiplicity and jet mass on $10{,}000$ test jets.

\begin{table}[t]
    \centering
    \caption{Full model and circuit-only AUC in three jet-multiplicity
    bins and three jet-mass bins, evaluated on $10{,}000$ test jets.
    Retention is the ratio of the circuit AUC to the full model AUC
    in the same bin. The circuit retains at least $90\%$ of the full
    model performance in every regime tested.}
    \label{tab:regimes}
    \begin{tabular}{lccc}
        \toprule
        Regime & Full AUC & Circuit AUC & Retention \\
        \midrule
        Sparse jets (few particles)     & $0.997$ & $0.991$ & $99.4\%$ \\
        Medium multiplicity              & $0.978$ & $0.960$ & $98.2\%$ \\
        Dense jets (many particles)      & $0.943$ & $0.873$ & $92.6\%$ \\
        \midrule
        Jet mass (low third)             & $0.948$ & $0.858$ & $90.6\%$ \\
        Jet mass (middle third)          & $0.950$ & $0.870$ & $91.6\%$ \\
        Jet mass (high third)            & $0.921$ & $0.850$ & $92.4\%$ \\
        \bottomrule
    \end{tabular}
\end{table}

The circuit retains $90.6$--$99.4\%$ of full-model performance across all
six regimes. The largest degradation occurs for dense jets ($92.6\%$
retention) and low-mass jets ($90.6\%$), both of which are kinematically
challenging regimes in which the full model itself has a smaller
advantage over classical methods. The high retention in sparse jets
($99.4\%$) is consistent with the expectation that the circuit, which
processes pairwise interactions through six attention heads, operates
most effectively when the relevant pairs are not obscured by combinatorial
background. The relatively uniform retention across the three jet-mass
bins ($90.6$--$92.4\%$) indicates that the circuit is not biased toward
any particular mass range.

%% file: sections/physical_content.tex
% ─────────────────────────────────────────────────────────────────────────────
\section{Physical Content of the Residual Stream}
\label{sec:representations}
% ─────────────────────────────────────────────────────────────────────────────
The circuit analysis of Section~\ref{sec:circuit} identifies which attention heads carry the model's discriminating computation but does not say what physical information those heads construct in the residual stream. In this section we address that question directly, by training linear probes that predict classical jet-substructure observables from the mean-pooled residual stream at each representation depth.

\subsection{Linear probing of jet-level observables}
\label{subsec:linearprobes}

We characterize the physical information encoded in the residual stream
at each layer by training Ridge regression linear probes to predict
jet-level observables from the mean-pooled particle representation. Each
probe is trained on $7{,}000$ jets and evaluated on $3{,}000$ held-out
jets; features are standardized before regression and the regularization
strength is set to $\alpha = 1.0$.

\begin{figure}[t]
    \centering
    \includegraphics[width=0.7\textwidth]{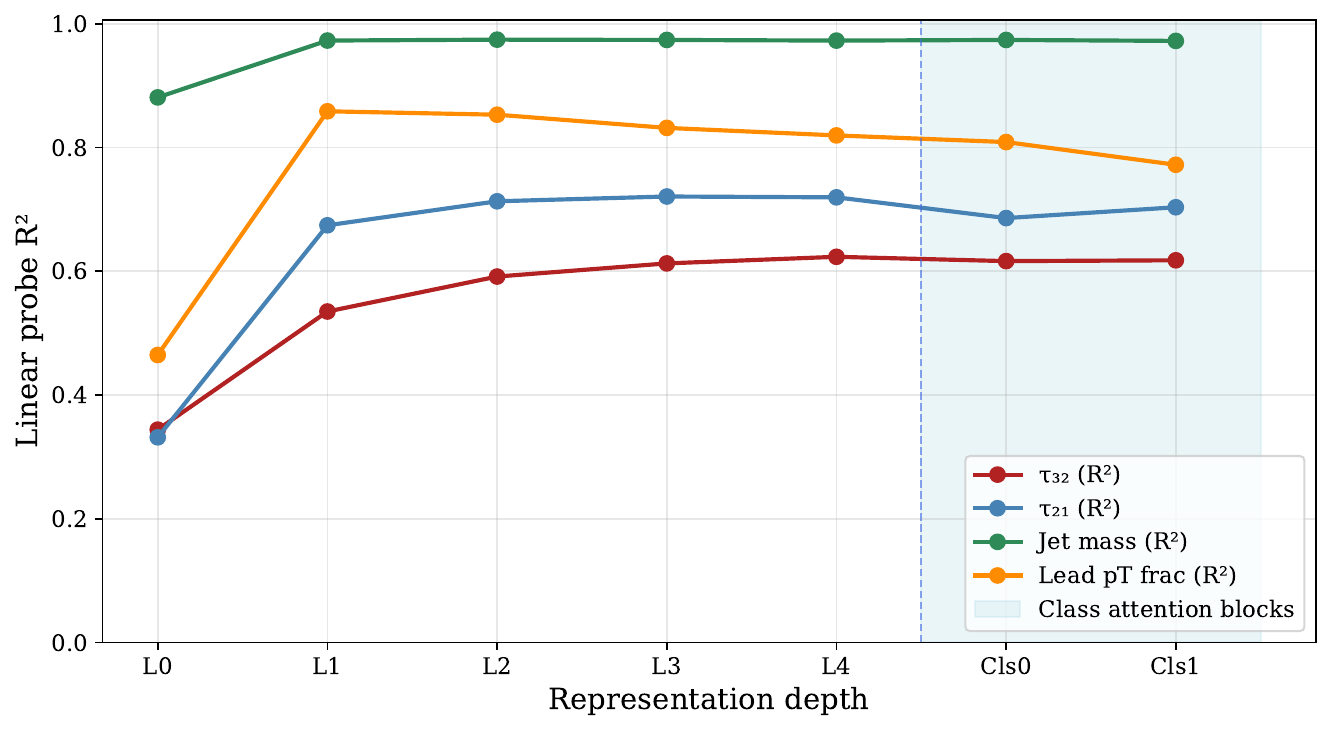}
    \caption{Linear probe $R^2$ for jet-level observables as a function
    of representation depth, including both particle attention layers
    (L0--L4) and class attention blocks (Cls0, Cls1). Jet mass and the
    leading-$p_T$ fraction are encoded with high $R^2$ from the
    embedding layer; $\tau_{32}$ and $\tau_{21}$ improve more
    gradually through the particle attention stack. The binary
    $\mathrm{is\_top}$ probe shows only modest improvement between
    L4 and the class attention blocks, consistent with the basis-rotation
    interpretation of Section~\ref{subsec:basisrotation}.}
    \label{fig:clsprobes}
\end{figure}

Figure~\ref{fig:clsprobes} reports the probe $R^2$ as a function of
depth. The probe results reveal a clear hierarchy. Jet mass is encoded
with $R^2 = 0.881$ at the embedding layer L0 and reaches $R^2 = 0.974$
by L1, remaining flat thereafter; this is consistent with jet mass
being a near-additive function of the per-particle features. The
leading-$p_T$ fraction follows a similar pattern, reaching $R^2 = 0.858$
at L1.

The $N$-subjettiness ratios require more processing. $\tau_{21}$
rises from $0.332$ at L0 to $0.713$ at L2; $\tau_{32}$ rises more
slowly, from $0.342$ at L0 to $0.660$ at L4. Neither $\tau_{32}$ nor
$\tau_{21}$ saturates within the particle attention stack at the
linear-probe level, consistent with these observables requiring
multi-particle reasoning that the attention mechanism progressively
constructs.

The binary $\mathrm{is\_top}$ probe reaches $R^2 = 0.868$ already at
L0 and rises to $R^2 = 0.929$ at Cls0, with no further improvement at
Cls1 ($0.927$). The flatness of the $\mathrm{is\_top}$ trajectory across
the particle attention layers, when read together with the basis-rotation
analysis in Section~\ref{subsec:basisrotation}, supports the same
mechanistic picture: the class signal is fully present in the residual
stream by L1 in some basis, and the class attention block aligns it
with the head's basis without introducing significant new information.

\subsection{Energy correlator probes: 2-prong versus 3-prong encoding}
\label{subsec:ecfprobes}

The classical energy correlation function family contains observables
designed for substructure of different prong multiplicities. The
two-point correlator $C_1^{(\beta)}$ is most sensitive to single-prong
versus diffuse radiation, the double ratio $C_2^{(\beta)}$ and the
two-prong-optimal observable $D_2^{(\beta)}$, given in 
Eq.~(\ref{eq:c2d2}), target two-prong substructure such as a hadronic
$W$ or $Z$ decay, and the next-order observables \(C_3^{(\beta)}\) and 
\( N_3^{(\beta)} \), as described in Appendix~\ref{subsec:jetsubstructure} by Eq.~\ref{eq:c3} and Eq.~\ref{eq:n3} respectively,
% \begin{equation}
%     C_3^{(\beta)}
%     \;=\;
%     \frac{\mathrm{ECF}(4,\beta)\, \mathrm{ECF}(2,\beta)}
%          {\mathrm{ECF}(3,\beta)^{2}}
%     \label{eq:c3}
% \end{equation}
% and the generalized correlator
% \begin{equation}
%     N_3^{(\beta)}
%     \;=\;
%     \frac{{}_{2}e_{4}^{(\beta)}}{\bigl({}_{1}e_{3}^{(\beta)}\bigr)^{2}}
%     \label{eq:n3}
% \end{equation}
target three-prong substructure such as the full $t \to W b \to
q\bar{q}b$ decay topology~\cite{Moult:2016cvt}. In Eq.~(\ref{eq:n3}),
${}_{v}e_{N}^{(\beta)}$ denotes the generalized correlator that uses
the $v$ smallest pairwise angles among the $\binom{N}{2}$ pairs in each
$N$-particle subset, normalized by the energy fractions $z_i = p_{T,i}
/ \mathrm{ECF}(1,\beta)$. By construction $C_3^{(\beta)}$ and
$N_3^{(\beta)}$ are the canonical IRC-safe observables for a 3-prong
tagging task: they are sensitive to the full three-body kinematics of
the top decay in a way different from the 2-prong observables.

We compute all the above observables for $\beta \in \{1, 2\}$ on
$5{,}000$ test jets, with $C_3^{(\beta)}$ and $N_3^{(\beta)}$ evaluated
on the leading $50$ and $40$ particles per jet respectively for
computational tractability ($\mathrm{ECF}(4)$ and ${}_{2}e_{4}$ are
$\mathcal{O}(n^{4})$ in the constituent count). Linear probe $R^{2}$
values are reported in Table~\ref{tab:ecfprobes} for the full set of
energy correlator and reference observables.

\begin{table}[t]
    \centering
    \caption{Linear probe $R^2$ for the energy correlator observables and
    classical reference observables at each representation depth. The
    peak $R^2$ across layers is given in the rightmost column, with
    the layer at which the peak is achieved indicated. The 3-prong
    observables $C_3^{(\beta=1)}$ and $N_3^{(\beta=1)}$, despite being
    the canonical targets for a 3-prong tagging task, are encoded
    less strongly than the corresponding 2-prong observables
    $C_2^{(\beta=1)}$ and $D_2^{(\beta=1)}$.}
    \label{tab:ecfprobes}
    \begin{tabular}{lccccccccc}
        \toprule
        Observable & L0 & L1 & L2 & L3 & L4 & Cls0 & Cls1 &
        Peak $R^2$ & Peak layer \\
        \midrule
        \multicolumn{10}{l}{\textit{2-prong energy correlators:}} \\
        $C_1^{(\beta=1)}$ & $0.856$ & $0.986$ & $0.989$ & $0.988$ &
            $0.989$ & $0.987$ & $0.986$ & $0.989$ & L2 \\
        $C_2^{(\beta=1)}$ & $0.670$ & $0.915$ & $0.930$ & $0.941$ &
            $0.945$ & $0.936$ & $0.935$ & $0.945$ & L4 \\
        $D_2^{(\beta=1)}$ & $0.639$ & $0.839$ & $0.842$ & $0.817$ &
            $0.804$ & $0.845$ & $0.837$ & $0.845$ & Cls0 \\
        \midrule
        \multicolumn{10}{l}{\textit{3-prong energy correlators:}} \\
        $C_3^{(\beta=1)}$ & $0.303$ & $0.648$ & $0.672$ & $0.662$ &
            $0.623$ & $0.639$ & $0.594$ & $0.672$ & L2 \\
        $N_3^{(\beta=1)}$ & $0.374$ & $0.756$ & $0.804$ & $0.790$ &
            $0.761$ & $0.772$ & $0.751$ & $0.804$ & L2 \\
        \midrule
        \multicolumn{10}{l}{\textit{$N$-subjettiness reference:}} \\
        $\tau_{21}$       & $0.189$ & $0.597$ & $0.663$ & $0.655$ &
            $0.672$ & $0.641$ & $0.646$ & $0.672$ & L4 \\
        $\tau_{32}$       & $0.310$ & $0.465$ & $0.544$ & $0.574$ &
            $0.581$ & $0.578$ & $0.571$ & $0.581$ & L4 \\
        \midrule
        Jet mass          & $0.862$ & $0.969$ & $0.972$ & $0.971$ &
            $0.971$ & $0.971$ & $0.969$ & $0.972$ & L2 \\
        \bottomrule
    \end{tabular}
\end{table}

\begin{figure}[t]
    \centering
    \includegraphics[width=\textwidth]{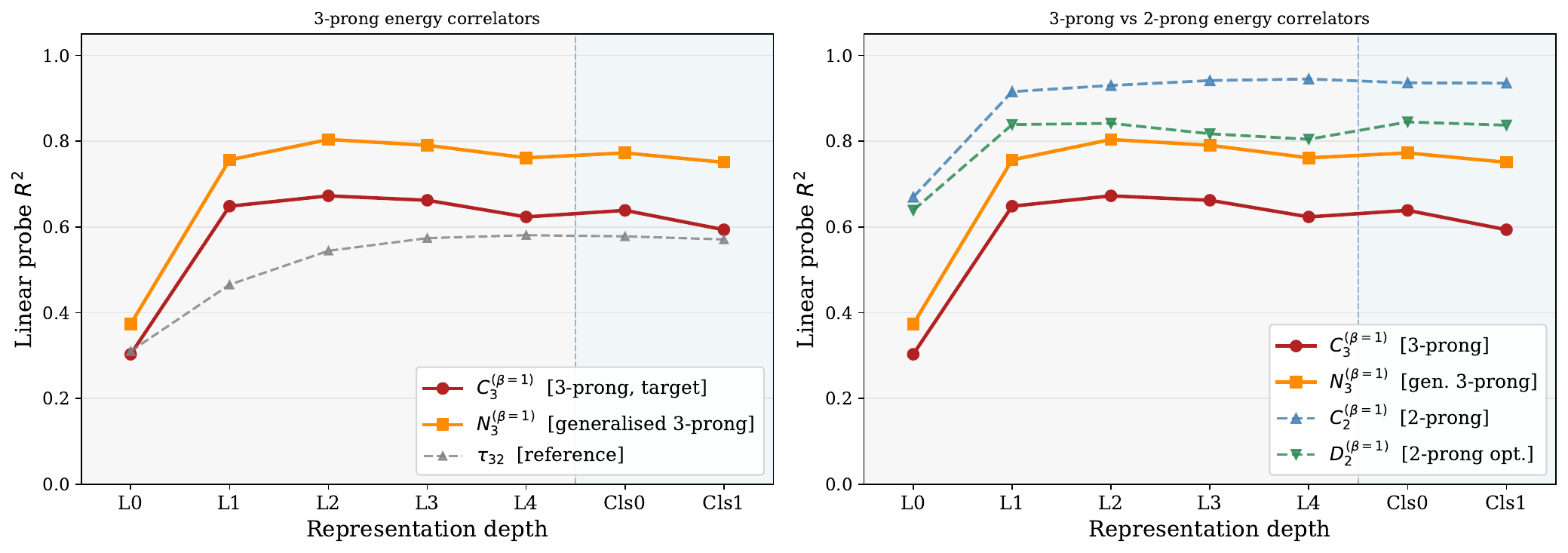}
    \caption{Linear probe $R^2$ for energy correlator observables as a
    function of representation depth. \emph{Left:} the 3-prong
    observables $C_3^{(\beta=1)}$ (red) and $N_3^{(\beta=1)}$ (orange)
    against the $N$-subjettiness reference $\tau_{32}$ (grey, dashed).
    \emph{Right:} the same 3-prong observables compared to the
    2-prong observables $C_2^{(\beta=1)}$ (blue, dashed) and
    $D_2^{(\beta=1)}$ (green, dashed). The shaded region marks the
    class attention blocks. The 3-prong observables are encoded
    significantly more strongly than $\tau_{32}$ but significantly
    less strongly than the corresponding 2-prong observables, despite
    being the canonical targets for the 3-prong tagging task.}
    \label{fig:ecf3prong}
\end{figure}

The probe values exhibit a clean and systematic ordering. Within the
energy correlator family, the 2-prong observables are encoded more
strongly than the 3-prong observables: the peak $R^2$ for
$C_2^{(\beta=1)}$ is $0.945$, compared to $0.804$ for
$N_3^{(\beta=1)}$ and $0.672$ for $C_3^{(\beta=1)}$. Within the 3-prong
group, the generalized correlator $N_3^{(\beta=1)}$ is encoded more
strongly than $C_3^{(\beta=1)}$, consistent with $N_3$ being more
robust to soft contamination because it uses only the two smallest
pairwise angles in each 4-particle subset. Both 3-prong observables
are encoded more strongly than the corresponding $N$-subjettiness
ratio $\tau_{32}$ (peak $R^2 = 0.581$).

\paragraph{The model encodes 2-prong structure more strongly than
3-prong structure.}
The most informative comparison is between the 3-prong observables
$C_3$ and $N_3$, which are the formally correct targets for a top
tagger, and the 2-prong observables $C_2$ and $D_2$, which target the
substructure of a single hadronic two-body decay. Despite being
trained on a 3-prong classification task, the model's residual stream
encodes 2-prong observables more strongly than 3-prong observables at
every layer beyond L0. This is not a small effect: at the peak layer
the gap is $|\Delta R^2| \approx 0.14$ between $C_2$ and $N_3$, and
$|\Delta R^2| \approx 0.17$ between $D_2$ and $C_3$.

We interpret this phenomenon as the model implicitly factorising the
top-tagging task into a more accessible sub-problem. In the kinematic
regime studied here, $p_T \in [550, 650]$~GeV, the top decay products
are collimated within an angular cone of $\Delta R \sim 2 m_t / p_T
\approx 0.5$, comfortably inside the $R = 0.8$ jet, and the visible
substructure of the top jet contains a hard, heavy two-prong
sub-system (the hadronic $W$ decay, $W \to q\bar{q}$, with $m_W \approx
80$~GeV) recoiling against a $b$-quark sub-jet. A 2-prong tagger
applied to such a jet identifies the $W$ system through its heavy
invariant mass; the full three-body topology, although in principle
more discriminating, is more demanding on the relative angular
arrangement of all three sub-jets and is more sensitive to soft
contamination. The model has converged on the easier of the two
available tagging strategies. The relay heads in the
identified circuit (Section~\ref{subsec:pathpatching}) attend
selectively to high-invariant-mass particle pairs
(Section~\ref{subsec:featattrib}) with positive class-discriminating
power on the pairwise ECF($2$) contribution
(Section~\ref{subsec:ecfattn}); these are precisely the kinematic
features of the heaviest single pair in the jet, which in a top jet
is dominated by the $W$-decay products. The 2-prong-over-3-prong
encoding ordering is the global linear-probe signature of this
local attentional bias.

\paragraph{Energy correlator basis preferred over $N$-subjettiness.}
A complementary comparison is between the energy correlator basis and
the $N$-subjettiness basis at fixed prong-count target. For 3-prong
substructure, $N_3^{(\beta=1)}$ ($R^2 = 0.804$) is encoded
substantially more strongly than $\tau_{32}$ ($R^2 = 0.581$); for
2-prong substructure, $C_2^{(\beta=1)}$ ($R^2 = 0.945$) is encoded
substantially more strongly than $\tau_{21}$ ($R^2 = 0.672$). The
energy correlator basis, which is constructed from pairwise (and
higher) products of $p_T$ and $\Delta R$, is more closely aligned with
the pairwise structure of the Particle Transformer's interaction bias
than the axis-based $N$-subjettiness construction. The model has
discovered a representation that is closer to the energy correlator
basis without either basis being explicitly encoded in the training
signal.

\paragraph{Mass-residualization control.}
A potential concern is that the apparent encoding advantages above
are driven by jet mass, which is itself encoded with $R^2 = 0.972$
already at L1 (Table~\ref{tab:ecfprobes}). Of the observables
discussed here, $D_2$ is the most strongly mass-correlated by
construction: a coefficient of determination of $0.59$ is obtained
when $D_2$ is regressed on jet mass alone, compared to $0.26$ for
$\tau_{32}$. To rule out mass as the driver of the
$D_2$-over-$\tau_{32}$ encoding gap, we residualize both observables
against jet mass via a degree-$3$ polynomial regression fitted on the
training subset and probe the residuals; the results are reported in
Table~\ref{tab:residualisation}. After residualization, the peak
$R^{2}$ for $D_2^{(\beta=1)}$ is $0.807$ and for $\tau_{32}$ is
$0.531$; the encoding advantage of $D_2$ over $\tau_{32}$ widens from
$0.243$ raw to $0.276$ after the contribution from jet mass has been
removed. The energy-correlator basis preference is therefore not a
mass artefact.

\begin{table}[t]
    \centering
    \caption{Linear probe $R^2$ for $D_2^{(\beta=1)}$ and $\tau_{32}$
    in their raw form and after residualization against jet mass. This test uses 7k train and 3k test jets, larger than for Table \ref{tab:ecfprobes} mainly because of cheaper computation requirements.
    Mass alone explains a coefficient of determination of $0.59$ for
    $D_2$ and $0.26$ for $\tau_{32}$; both observables retain
    substantial mass-independent information. After residualization,
    the $D_2$ encoding remains well above the $\tau_{32}$ encoding at
    every layer, confirming that the model's preferential encoding of
    $D_2$ is not a mass artefact.}
    \label{tab:residualisation}
    \begin{tabular}{lccccccccc}
        \toprule
        Observable & L0 & L1 & L2 & L3 & L4 & Cls0 & Cls1 & Peak $R^2$ \\
        \midrule
        $D_2^{(\beta=1)}$ raw                & $0.741$ & $0.880$ & $0.894$ & $0.889$ & $0.883$ & $0.901$ & $0.903$ & $0.903$ \\
        $D_2^{(\beta=1)}$ mass-residualized  & $0.368$ & $0.767$ & $0.796$ & $0.790$ & $0.785$ & $0.800$ & $0.807$ & $0.807$ \\
        $\tau_{32}$ raw                       & $0.342$ & $0.561$ & $0.600$ & $0.633$ & $0.660$ & $0.650$ & $0.645$ & $0.660$ \\
        $\tau_{32}$ mass-residualized         & $0.141$ & $0.397$ & $0.454$ & $0.496$ & $0.531$ & $0.518$ & $0.515$ & $0.531$ \\
        \bottomrule
    \end{tabular}
\end{table}

\begin{figure}[t]
    \centering
    \includegraphics[width=\textwidth]{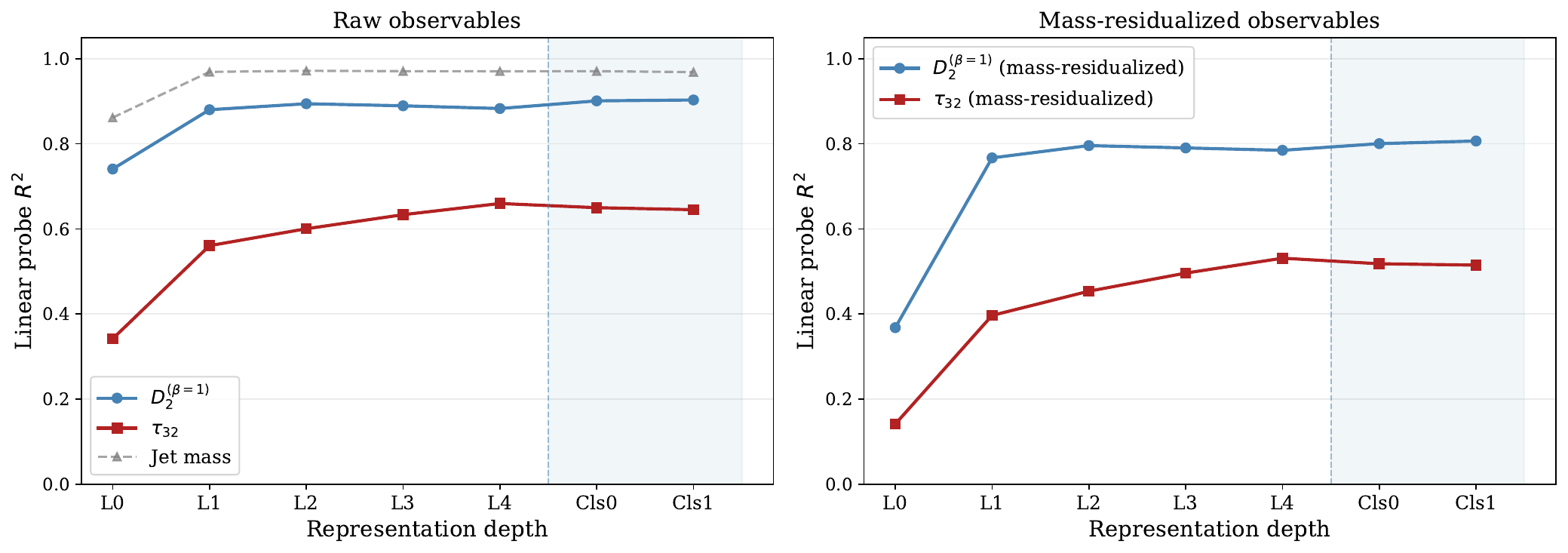}
    \caption{Linear probe $R^2$ for $D_2^{(\beta=1)}$ (blue) and
    $\tau_{32}$ (red) in raw form (left) and after residualization
    against jet mass (right). The dashed grey line on the left panel
    shows the jet mass probe $R^2$ for reference. The
    $D_2$-over-$\tau_{32}$ encoding advantage is preserved, and indeed
    widens slightly, after the contribution from jet mass has been
    removed.}
    \label{fig:residualised}
\end{figure}

%% file: sections/features_sec.tex
% ─────────────────────────────────────────────────────────────────────────────
\section{Interaction Features Driving Each Head}
\label{sec:features}
% ─────────────────────────────────────────────────────────────────────────────

\subsection{Correlation between attention and pairwise features}
\label{subsec:featattrib}

The attention bias $U^{(h)}_{ij}$ in Eq.~(\ref{eq:attention}) is
computed by the pairwise MLP from the four pairwise features $\ln k_T$,
$\ln z$, $\ln \Delta$, and $\ln m^2$. To understand which feature each
head responds to, we measure the Pearson correlation between the
attention weight $A^{(l,h)}_{ij}$ and each of the four features over
all off-diagonal real particle pairs across $2{,}000$ test jets.
Table~\ref{tab:featattrib} reports the result for the four heads with
the most substantial direct effects (excluding the readout head L3H3
and the Layer~$1$ head L1H1, whose ECF-attention correlations are
discussed in Section~\ref{subsec:ecfattn}).

\begin{table}[h]
    \centering
    \caption{Pearson correlation between attention weights
    $A^{(l,h)}_{ij}$ and the four pairwise interaction features for
    four representative circuit heads. Correlations are pooled over
    all off-diagonal real particle pairs in $2{,}000$ test jets.
    L0H1 stands apart with negative correlations on $\ln k_T$,
    $\ln \Delta$, and $\ln m^2$; the other three heads show positive
    correlations on $\ln k_T$ and $\ln m^2$ and negative correlations
    on $\ln z$.}
    \label{tab:featattrib}
    \begin{tabular}{lcccc}
        \toprule
        Head & $\ln k_T$ & $\ln z$ & $\ln \Delta$ & $\ln m^2$ \\
        \midrule
        L0H1 & $-0.302$ & $+0.152$ & $-0.481$ & $-0.429$ \\
        L0H2 & $+0.384$ & $-0.152$ & $+0.185$ & $+0.504$ \\
        L1H0 & $+0.250$ & $-0.308$ & $+0.108$ & $+0.439$ \\
        L1H3 & $+0.315$ & $-0.201$ & $+0.141$ & $+0.454$ \\
        \bottomrule
    \end{tabular}
\end{table}

L0H1 stands apart with negative correlations on $\ln k_T$ ($-0.302$),
$\ln \Delta$ ($-0.481$), and $\ln m^2$ ($-0.429$): it preferentially
attends to pairs with small angular separation, low transverse-momentum
product, and low pairwise invariant mass. This is the kinematic regime
of soft and collinear emissions, not of hard substructure. We return to
the interpretation of this pattern in
Section~\ref{subsec:algorithm}.

The other three heads (L0H2, L1H0, L1H3) show a common pattern of
positive correlations with $\ln m^2$ ($+0.44$ to $+0.50$) and $\ln k_T$
($+0.25$ to $+0.38$) and negative correlations with $\ln z$ ($-0.15$ to
$-0.31$). These heads preferentially attend to particle pairs with large
invariant mass, high $k_T$, and asymmetric momentum sharing, which are
precisely the kinematic properties of decay products from a heavy
particle. The hadronic decay products of the $W$ boson in
$t \to W b \to q\bar{q}b$ provide the dominant source of large-invariant-mass,
high-$k_T$ pairs in a top jet, and the relay heads attention pattern is
consistent with localizing on these pairs.

\subsection{Causal feature ablation}
\label{subsec:featablation}

The Pearson correlations of Section~\ref{subsec:featattrib} are
correlational. To establish causal dependence, we zero each pairwise
feature individually in the input to the pairwise MLP, re-run the model,
and measure the resulting change in the logit difference and in the
attention pattern.

\begin{figure}[t]
    \centering
    \includegraphics[width=\textwidth]{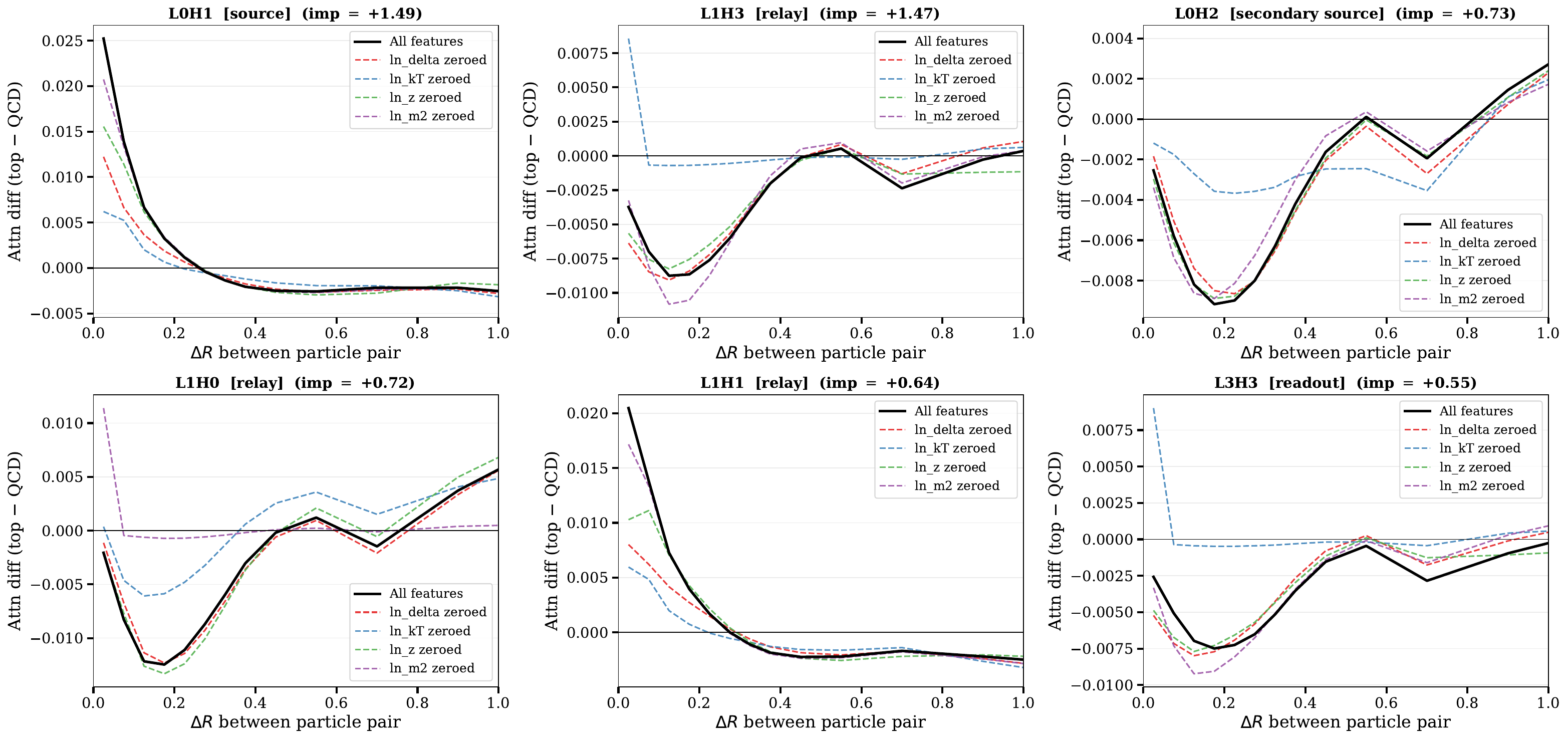}
    \caption{Causal feature ablation. For each of the four pairwise
    features (coloured dashed lines), the feature is zeroed in the
    input tensor before the pairwise MLP and the model is re-run.
    The plotted quantity is the difference in the head's mean
    attention weight between top and QCD jets as a function of
    pairwise $\Delta R$. A large deviation from the baseline (black)
    indicates that the head causally depends on that feature.
    Profiles are averaged over $10{,}000$ test jets. The six panels
    correspond to six heads spanning a range of zero-ablation
    importances, the value of which is given in each panel title.}
    \label{fig:featablation}
\end{figure}

Figure~\ref{fig:featablation} shows the effect of each single-feature
ablation on the top-versus-QCD attention difference profile of four
heads. Zeroing $\ln k_T$ produces the largest change in the mean logit
difference across all single-feature ablations ($+4.44$ when zeroed,
where the convention is that a positive change is a degradation),
identifying $\ln k_T$ as the most causally important pairwise feature
for the model's overall performance. The relative profile changes are
substantial for all four heads (relative $L^2$ change in the $\Delta R$-resolved
attention difference: $0.86$--$1.44$), consistent with $\ln k_T$ being
a primary driver of attention patterns across heads.

The ablation also reveals head-specific dependencies. Zeroing $\ln m^2$
produces a disproportionately large profile change in the relay-type
panels relative to the source-type panel, consistent with the
correlation pattern in Table~\ref{tab:featattrib} showing that the relay
heads have stronger $\ln m^2$ correlations than L0H1.

A counter-intuitive observation is that zeroing $\ln z$ produces a
\emph{positive} change in the mean logit difference ($+2.69$), meaning
that the model performs slightly better when this feature is removed.
A possible interpretation can be that $\ln z$, which measures the momentum sharing
asymmetry of a pair, can drive the attention onto soft fragments with
high $z$ asymmetry but small absolute momentum scale, and removing this
input partially suppresses a confounding signal. We do not interpret
this as a sign that $\ln z$ should be removed from the architecture in
practice; the effect is small compared to the $\ln k_T$ effect, and
more importantly the $\ln z$ feature is a non-trivial component of the
representation underlying the relay heads' positive direct effects on
$\ln m^2$, which would also be affected by its removal in non-trivial
ways.

\subsection{ECF-attention correlation and class-discriminating attention}
\label{subsec:ecfattn}

The interaction-feature correlations of Section~\ref{subsec:featattrib}
characterize each head's attention pattern at the level of single
pairwise features ($\ln k_T$, $\ln z$, $\ln \Delta$, $\ln m^{2}$). A
complementary characterization, more directly tied to the
substructure observables of Section~\ref{subsec:ecfprobes}, asks
whether the head attends selectively to particle pairs that
participate in higher-order energy correlator structure, and whether
this selective attention is class-dependent.

For each pair $(i,j)$ in a jet we construct two scalar proxies that
quantify the contribution of that pair to substructure of a given
prong-count. The 2-prong proxy is the (normalized) pairwise
contribution to $\mathrm{ECF}(2,\beta)$,
\begin{equation}
    \tilde{c}_{2;\,ij}
    \;=\;
    \frac{p_{T,i}\, p_{T,j}\, \Delta R_{ij}^{\beta}}
         {\mathrm{ECF}(1)^{2}},
    \label{eq:c2_proxy}
\end{equation}
which depends only on the kinematics of the pair $(i,j)$ itself. The
3-prong proxy is the pair's marginal contribution to
$\mathrm{ECF}(3,\beta)$, obtained by summing the energy correlator
weight over the choice of third particle $k$,
\begin{equation}
    \tilde{c}_{3;\,ij}
    \;=\;
    \frac{p_{T,i}\, p_{T,j}\, \Delta R_{ij}^{\beta}}
         {\mathrm{ECF}(1)^{3}}
    \sum_{k \neq i,j} p_{T,k}\, \Delta R_{ik}^{\beta}\, \Delta R_{jk}^{\beta}.
    \label{eq:c3_proxy}
\end{equation}
By construction, $\tilde{c}_{3;\,ij}$ is large when the pair $(i,j)$
sits at the apex of a 3-particle configuration with a third high-$p_T$
particle at small angular separation from both $i$ and $j$. In a top
jet, the dominant such configurations are those involving the three
hard sub-jets of the $t \to W b \to q\bar{q}b$ decay; in a QCD jet,
they are produced through ordinary parton-shower evolution. The two
proxies $\tilde{c}_{2;\,ij}$ and $\tilde{c}_{3;\,ij}$ are not
mutually exclusive: every pair has both, and the proxies factorize the
contribution of pair $(i,j)$ to the energy correlator structure of
the jet by prong-count.

For each circuit head $(l,h)$ and each prong-count $n \in \{2, 3\}$
we compute the Pearson correlation $r_n(l,h)$ between the head's
attention weight $A^{(l,h)}_{ij}$ and the corresponding proxy
$\tilde{c}_{n;\,ij}$, separately for top jets and for QCD jets. The
class-discriminating power of the head on the $n$-prong proxy is
then defined as
\begin{equation}
    \delta_n(l,h)
    \;\equiv\;
    r_n^{\mathrm{top}}(l,h) \,-\, r_n^{\mathrm{QCD}}(l,h).
\end{equation}
A positive $\delta_n$ indicates that the head's attention is more
selective for $n$-prong-relevant pairs in top jets than in QCD jets;
$\delta_n \approx 0$ indicates that the head's attention is class-agnostic at
the level of $n$-prong pairwise contributions. The pair $(\delta_2,
\delta_3)$ characterizes whether the head's class-discriminating
attention operates predominantly at the 2-particle or the 3-particle
level.

\begin{figure}[t]
    \centering
    \includegraphics[width=0.7\textwidth]{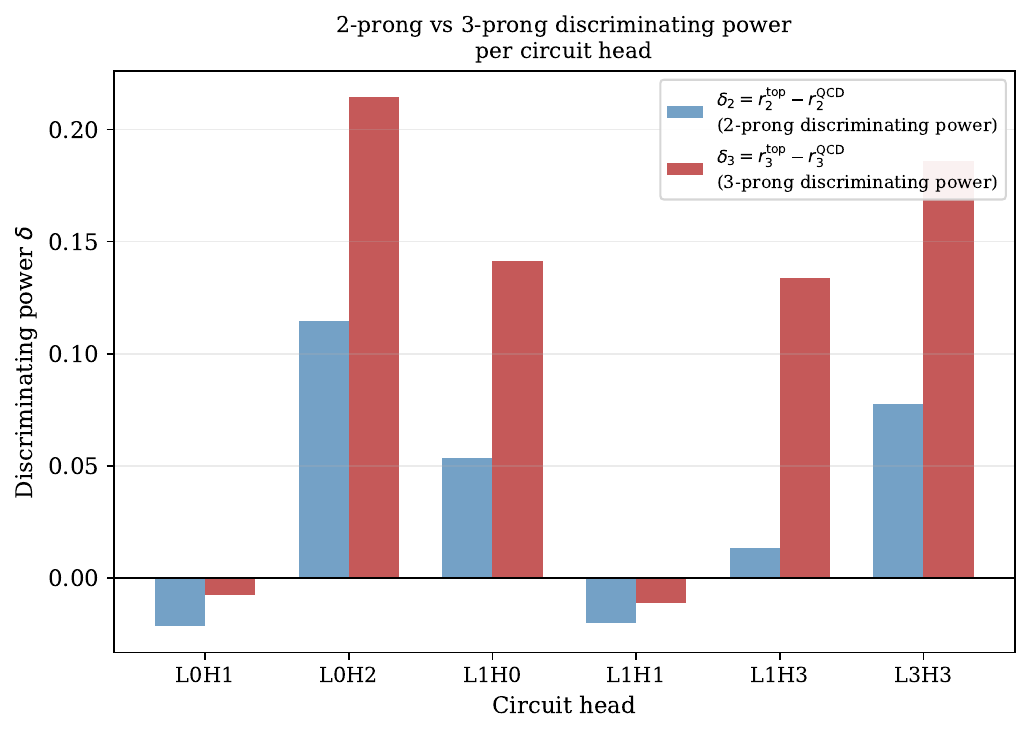}
    \caption{Class-discriminating powers $\delta_2$ (blue) and
    $\delta_3$ (red) for the six circuit heads, computed from the
    Pearson correlations between the head's attention weights and the
    pairwise proxies of Eqs.~(\ref{eq:c2_proxy})--(\ref{eq:c3_proxy})
    on top and QCD jets over $2{,}000$ test jets. For L0H2, L1H0,
    L1H3, and L3H3, the 3-prong discriminating power exceeds the
    2-prong discriminating power; for L0H1 and L1H1, both are
    negligible.}
    \label{fig:ecfattn}
\end{figure}

Table~\ref{tab:ecfattn} reports both proxies for the six circuit
heads, computed over $2{,}000$ test jets;
Figure~\ref{fig:ecfattn} displays the same data graphically.

\begin{table}[t]
    \centering
    \caption{Pearson correlations between attention weights and the
    2-prong and 3-prong pairwise proxies of
    Eqs.~(\ref{eq:c2_proxy})--(\ref{eq:c3_proxy}), computed
    separately on top and QCD jets over $2{,}000$ test jets. The
    class-discriminating powers $\delta_n = r_n^{\mathrm{top}} -
    r_n^{\mathrm{QCD}}$ measure how selectively each head attends to
    $n$-prong pairwise contributions in top jets relative to QCD
    jets. For four of the six circuit heads (L0H2, L1H0, L1H3, L3H3),
    the 3-prong discriminating power $\delta_3$ exceeds the 2-prong
    discriminating power $\delta_2$.}
    \label{tab:ecfattn}
    \begin{tabular}{lcccccc}
        \toprule
        Head & $r_2^{\mathrm{top}}$ & $r_2^{\mathrm{QCD}}$ & $\delta_2$
             & $r_3^{\mathrm{top}}$ & $r_3^{\mathrm{QCD}}$ & $\delta_3$ \\
        \midrule
        L0H1 & $-0.092$ & $-0.071$ & $-0.021$ & $-0.103$ & $-0.095$ & $-0.007$ \\
        L0H2 & $+0.538$ & $+0.423$ & $+0.114$ & $+0.493$ & $+0.279$ & $+0.214$ \\
        L1H0 & $+0.373$ & $+0.320$ & $+0.054$ & $+0.370$ & $+0.228$ & $+0.141$ \\
        L1H1 & $-0.101$ & $-0.081$ & $-0.020$ & $-0.114$ & $-0.103$ & $-0.011$ \\
        L1H3 & $+0.511$ & $+0.498$ & $+0.014$ & $+0.470$ & $+0.337$ & $+0.134$ \\
        L3H3 & $+0.552$ & $+0.475$ & $+0.078$ & $+0.521$ & $+0.335$ & $+0.186$ \\
        \bottomrule
    \end{tabular}
\end{table}

The six circuit heads separate into two clear groups. The primary
source L0H1 and the relay head L1H1 have $\delta_2$ and $\delta_3$
both close to zero (and slightly negative), indicating that their
attention is class-agnostic at the level of pairwise correlator
contributions. This is consistent with the interaction-feature
analysis of Section~\ref{subsec:featattrib}: L0H1's attention is
biased toward soft and collinear pairs, which are not the pairs that
contribute to either the 2-prong or the 3-prong proxy in a
class-discriminating way.

The remaining four heads, the secondary source L0H2, the relay
heads L1H0 and L1H3, and the readout L3H3, all show positive
class-discriminating power on both proxies, with the same systematic
ordering: $\delta_3 > \delta_2$ in all four cases, with $\delta_3$
exceeding $\delta_2$ by factors of $1.9$ (L0H2), $2.6$ (L1H0), $9.6$
(L1H3), and $2.4$ (L3H3). The relay head L1H3 is particularly
striking: it is class-agnostic on the 2-prong proxy ($\delta_2 =
0.014$) but has a sizeable class-discriminating power on the 3-prong
proxy ($\delta_3 = 0.134$), indicating that its class-selective
attention operates almost entirely at the 3-particle level.

This $\delta_3 > \delta_2$ hierarchy is a non-trivial finding when
read alongside the linear probe analysis of
Section~\ref{subsec:ecfprobes}, which showed that the residual stream
encodes 2-prong observables more strongly than 3-prong observables
($C_2$ at $R^2 = 0.945$ versus $N_3$ at $R^2 = 0.804$, and $D_2$ at
$R^2 = 0.845$ versus $C_3$ at $R^2 = 0.672$). The two findings appear
in tension at first reading: how can the heads' attention be more
3-prong-selective than 2-prong-selective if the resulting residual
stream is more 2-prong-encoded than 3-prong-encoded?

The reconciliation is that the 3-prong selectivity lives in the
attention computation, while the residual stream receives the
per-particle write-out. The pairwise-bias mechanism in the Particle
Transformer (Section~\ref{subsec:part}) gives every attention head
access to all pairwise features simultaneously and allows the head to
attend selectively to pairs $(i,j)$ in a way that depends on the
presence of a third particle $k$ (i.e., on a 3-particle configuration),
because the bias $U^{(h)}_{ij}$ is constructed from features that
implicitly encode this dependence. The output that each head writes
to the residual stream, however, is a per-particle vector, and the
linear probes of Section~\ref{subsec:ecfprobes} read these
representations in mean-pooled (per-jet) form. The 3-prong selection
mechanism is therefore consumed in the attention computation but is
not preserved as an explicit 3-prong representation in the residual
stream; what is preserved is a 2-prong-dominant readout whose
amplitude is modulated by 3-prong-aware attention. The 2-prong
encoding seen by the linear probes and the 3-prong attention pattern
seen by the class-discriminating-power analysis are therefore two
sides of the same computation.

%% file: sections/discussion.tex
% ─────────────────────────────────────────────────────────────────────────────
\section{Discussion}
\label{sec:discussion}
% ─────────────────────────────────────────────────────────────────────────────
In this section we take a step ahead to understand the inter-connection between all the results that we systematically obtained in the previous sections. 

\subsection{The circuit as a physical algorithm}
\label{subsec:algorithm}

Combining the evidence from the preceding sections, the Particle
Transformer can be described as implementing top-quark tagging through a
two-stage algorithm with the following structure.

\paragraph{Stage 1: feature extraction in the residual stream.}
The four particle attention layers act jointly to construct a
near-complete classification representation in the residual stream by L1
(per-layer logistic probe AUC $0.971$ at L1, rising marginally to $0.977$
by L4; Section~\ref{subsec:basisrotation}). This representation encodes
two-prong energy correlator structure in particular: the genuine
mass-residualized $D_2^{(\beta=1)}$ encoding rises from $R^2 = 0.368$ at
L0 to $R^2 = 0.767$ at L1 (Section~\ref{subsec:ecfprobes}). Within this
representation, the identified six-head circuit handles the bulk of the
discriminating computation. Its components have qualitatively distinct
roles:

\begin{itemize}
    \item L0H1 (primary source) attends preferentially to soft and
    collinear particle pairs (negative $\ln k_T$, $\ln \Delta$, $\ln m^2$
    correlations; Section~\ref{subsec:featattrib}) and has a class-agnostic
    relationship to ECF($2$) at the level of pairwise contributions
    (Section~\ref{subsec:ecfattn}). Despite this, it has the largest
    direct effect ($+4.25$) and is by itself sufficient to recover
    $88.6\%$ of the full model AUC. We interpret L0H1 as performing a
    \emph{soft-radiation contextualization} step: by summarizing the soft
    and collinear emission pattern in a class-agnostic way, it provides
    the downstream relay heads with a normalization against which the
    hard substructure can be identified. The strong path effects from
    L0H1 to all three Layer~$1$ relay heads ($1.19$--$1.45$;
    Section~\ref{subsec:patheffects}) are consistent with this role.

    \item L0H2 (secondary source) is nearly parallel to L0H1 in
    representational space (cosine similarity $0.92$;
    Section~\ref{subsec:similarity}) but, in contrast to L0H1, has positive
    correlations with $\ln k_T$, $\ln m^2$ and a positive
    class-discriminating ECF($2$) attention pattern ($\delta = +0.112$;
    Section~\ref{subsec:ecfattn}). It plays the role of a
    class-discriminating complement to L0H1 in Layer~$0$, contributing
    weaker but still significant path effects to the same downstream
    targets ($0.85$--$0.90$).

    \item L1H0, L1H1, L1H3 (relay heads) all attend preferentially to
    high-invariant-mass, high-$k_T$, asymmetric-$z$ pairs
    (Section~\ref{subsec:featattrib}) and form a tight representational
    cluster (cosine similarities $0.81$--$0.95$). Their joint role is to
    localize the hard two-prong substructure characteristic of the
    $W$-boson decay $W \to q\bar{q}$. Their negative direct effects
    confirm that their clean-run activations are conditional on the
    upstream signal from the source heads; their large pairwise
    super-additivities confirm that they are partly redundant.

    \item L3H3 (readout) has low representational similarity to all
    other circuit heads ($\le 0.48$), a positive direct effect of $+0.71$,
    and a positive class-discriminating ECF($2$) attention
    ($\delta = +0.076$). It performs a final aggregation operation that
    combines the relayed signal into the format consumed by the class
    attention block.
\end{itemize}

\paragraph{Stage 2: basis alignment by the class attention block.}
The class attention block Cls0 produces the dramatic logit-lens jump
from $0.111$ to $0.973$ in a single step (Section~\ref{subsec:logitlens_results}).
The per-layer probe analysis resolves this jump as a basis rotation
(Section~\ref{subsec:basisrotation}): the linearly accessible class
information is essentially complete by L1 in the particle attention
basis, and Cls0 reorients this signal into the basis on which the
final classification head was trained. The second class attention block
Cls1 adds essentially no information ($+0.006$ logit-lens AUC), as
expected for a basis-aligned representation.

This two-stage picture is conceptually compact and resembles the
structure of circuits identified in language models in the IOI
literature~\cite{2022arXiv221100593W}, in which an explicit name-mover computation
is preceded by upstream backup and inhibition heads. The present work
establishes that the source-relay-readout architectural pattern arises
in a physics classifier trained on a physically motivated task without
language-model-style structure being built into the data, and is
therefore unlikely to be a feature specific to natural language.

\subsection{Connection to classical jet substructure observables}
\label{subsec:classicalconn}

The probe analysis of Section~\ref{subsec:ecfprobes} shows that the
model's residual stream encodes $D_2^{(\beta=1)}$ to peak $R^2 = 0.903$
in raw form and $0.807$ after residualization against jet mass, while
$\tau_{32}$ is encoded to $0.660$ raw and $0.531$ residualized. The
gap between $D_2$ and $\tau_{32}$ is not a consequence of the encoding
of jet mass and persists, indeed widens, after mass is removed.

A natural interpretation is that the model has implicitly discovered a
two-prong tagger as the dominant component of its top tagger. Within a
top jet, the dominant pairs that contribute to large-invariant-mass
configurations are the daughters of the hadronic $W$ decay, which has
$m_W \approx 80$~GeV. The model's relay heads attend selectively to
exactly these pairs (positive $\ln m^2$ correlations and positive
ECF($2$) class-discriminating power on L0H2, L1H0, L3H3;
Sections~\ref{subsec:featattrib} and~\ref{subsec:ecfattn}). The
two-prong-optimal observable $D_2$, which is constructed from
$\mathrm{ECF}(2)$ and $\mathrm{ECF}(3)$ to be optimal under IRC-safe
conditions for two-prong versus one-prong discrimination~\cite{Larkoski:2014gra},
matches this internal structure more closely than $\tau_{32}$, which is
constructed instead around the residual radiation about three-prong
axes. The model has organized its computation around the most
identifiable substructure, i.e. the two-prong $W$ decay, rather than
the full three-body topology of the top decay.

This is, to our knowledge, the first explicit demonstration of using causal circuit analysis for analysing jet-tagger interpretability.

\subsection{Limitations}
\label{subsec:limitations}

The analysis presented here has several limitations.

First, the model studied is a small Particle Transformer ($4$ particle
attention layers, $4$ heads per layer, $1.3$M parameters)
trained on the standard top-tagging benchmark. Larger and more capable
Particle Transformers, trained at higher AUC ($0.985$ and above), may
have richer circuit structures that are not captured by a six-head
analysis. The methods developed here scale directly, as in studies pertaining natural language, but the specific
six-head identification should not be assumed universal.

Second, the random-baseline comparison uses $200$ randomly sampled
six-head subsets out of $\binom{16}{6} = 8008$ possible subsets. This
sampling is sufficient to place the identified circuit at the $96$th
percentile of the random distribution with empirical standard deviation
$\sigma = 0.125$. However, a complete evaluation will provide the final word on this aspect.

Third, the linear probes used throughout the residual-stream analysis
measure linear accessibility of information, not the full information
content. Observables that are encoded non-linearly would be detected at
lower $R^2$ than they truly are, and the values reported here should be
read as lower bounds on the true encoding fidelity.

Fourth, the path-patching analysis uses two on-manifold corruption
strategies in within-batch replacement, jet permutation as
discussed in Section~\ref{subsec:corruption_robustness}, and we have
documented that off-manifold (Gaussian) corruption is structurally
incompatible with the standard recovery-score formulation on this
dataset. The qualitative sign pattern is robust within the on-manifold
class; quantitative direct-effect magnitudes vary by up to a factor of
two between strategies and should be interpreted with this caveat.

Fifth, although we have verified the cross-seed stability of the
zero-ablation importance ranking across five training seeds, the
detailed path-patching, probing, and feature-attribution analyses use
the seed-$0$ model. Extending the path-patching analysis to multiple
seeds would test whether the source-relay-readout sign pattern (rather
than the specific head identities) is universal. We expect this to be
the case based on the cross-seed stability of zero-ablation importance,
but we leave a full multi-seed circuit verification to future work.

%% file: sections/conclusion.tex
% ─────────────────────────────────────────────────────────────────────────────
\section{Conclusion}
\label{sec:conclusion}
% ─────────────────────────────────────────────────────────────────────────────

We have performed a complete mechanistic interpretability analysis of a
Particle Transformer trained for top quark tagging on the standard
benchmark dataset. Combining zero ablation across five training seeds,
path patching with two complementary on-manifold corruption strategies,
bootstrap-resolved path-effect estimates, linear probing of the residual
stream, and per-layer trained probes that resolve basis-rotation effects
in the standard logit lens, we have arrived at the following picture of
the model's internal computation.

A sparse six-head circuit recovers $97.3\%$ of the full model AUC and
lies at the $96$-th percentile of $200$ randomly sampled six-head subsets. The circuit has a source-relay-readout structure: the primary source L0H1 carries nearly all the causal discriminating information (direct effect
$+4.25$); the secondary source L0H2 contributes a weaker, complementary
signal; three Layer~$1$ relay heads localize high-invariant-mass particle
pairs and require the upstream context to function correctly; and the
single Layer~$3$ readout head L3H3 aggregates the relayed signal. The
sign pattern of the direct effects is robust under two on-manifold
corruption strategies, and we have documented and explained a structural
incompatibility between off-manifold (Gaussian) corruption and the
standard recovery-score formulation on this kinematically narrow dataset.

The residual stream encodes the two-prong-optimal energy correlator
$D_2^{(\beta=1)}$ to peak coefficient of determination $R^2 = 0.90$ in
raw form and $0.81$ after residualization against jet mass, both
substantially above the corresponding values for $N$-subjettiness
$\tau_{32}$ ($R^2 = 0.66$ and $0.53$ respectively). The model has
implicitly organized its computation around the two-prong substructure of
the hadronic $W$ decay, in agreement with the relay heads' attention
selectively to high-invariant-mass particle pairs.

Resolving the standard logit-lens analysis with per-layer trained
logistic probes, we have shown that the apparent commitment of the model
to a classification decision in the first class attention block is in
fact closer to a basis rotation: the linearly accessible class signal saturates to
$\mathrm{AUC} \approx 0.97$ already in the first particle attention layer,
and the class attention block aligns this signal with the basis on which
the trained classifier head operates.

These results demonstrate that mechanistic interpretability methods
developed in the context of natural language models transfer, with care,
to jet physics classifiers. They reveal that a Particle Transformer
trained on raw four-momenta can rediscover a physically meaningful
energy-correlator structure through gradient descent alone, and they
suggest that the source-relay-readout architecture observed here may be
characteristic of a broader class of physics-domain Transformers. Future
work should apply these methods to larger Particle Transformer variants
and to other classification tasks, test the universality of the
source-relay-readout architecture, and explore whether the
basis-rotation interpretation of the class attention block generalizes
beyond top tagging.

%% file: sections/acknowledgment.tex
% ─────────────────────────────────────────────────────────────────────────────
\section*{Acknowledgements}
% ─────────────────────────────────────────────────────────────────────────────
S.R. acknowledges the Ph.D fellowship from IIT-Kanpur. \\ S.G. is supported by the IIT-Kanpur faculty initiation grant (IITK /PHY /2023499) and Anusandhan National Research Foundation, Advanced Research Grant (ANRF/ARG/2025/005801/PS). \\The training of the models are done using SWAN GPU facility at CERN.  

%% file: sections/appendix.tex
\appendix

% ─────────────────────────────────────────────────────────────────────────────
\section{Appendix}
\label{sec:background}
% ─────────────────────────────────────────────────────────────────────────────

\subsection{Jet substructure for top quark tagging}
\label{subsec:jetsubstructure}

A boosted top quark with transverse momentum $p_T \in [550, 650]$~GeV decays
to a jet whose constituents encode the kinematics of the three-body final
state $q\bar{q}b$. Two classical families of observables are particularly
relevant to the present work.

\paragraph{N-subjettiness.}
The $N$-subjettiness observable~\cite{Thaler:2010cxa,Thaler:2011gf} is defined as
\begin{equation}
    \tau_N
    \;=\;
    \frac{1}{d_0}
    \sum_{k} p_{T,k}\,
    \min_{j}\, \Delta R(k, \hat{n}_j)^{\beta},
    \qquad
    d_0 \;=\; \sum_{k} p_{T,k}\, R_0^{\beta},
    \label{eq:tauN}
\end{equation}
where the sum runs over jet constituents, the minimum is taken over $N$
candidate subjet axes $\{\hat{n}_j\}$, $R_0$ is the jet radius, and $\beta$
is an angular weighting exponent. The ratios $\tau_{21} \equiv \tau_2 / \tau_1$
and $\tau_{32} \equiv \tau_3 / \tau_2$ are sensitive to the presence of
two-prong and three-prong substructure, respectively, and $\tau_{32}$ is
the canonical $N$-subjettiness variable for top tagging.

\paragraph{Energy correlation functions.}
The (non-normalized) energy correlation functions~\cite{Larkoski:2013eya} of
order $n$ and angular weight $\beta$ are defined by
\begin{align}
    \mathrm{ECF}(1,\beta) &= \sum_i p_{T,i}, \\
    \mathrm{ECF}(2,\beta) &= \sum_{i<j} p_{T,i}\, p_{T,j}\, \Delta R_{ij}^{\beta}, \\
    \mathrm{ECF}(3,\beta) &= \sum_{i<j<k} p_{T,i}\, p_{T,j}\, p_{T,k}\,
        (\Delta R_{ij}\, \Delta R_{ik}\, \Delta R_{jk})^{\beta}.
\end{align}
The double ratios
\begin{equation}
    C_2^{(\beta)}
    \;=\;
    \frac{\mathrm{ECF}(3,\beta)\, \mathrm{ECF}(1,\beta)}
         {\mathrm{ECF}(2,\beta)^{2}},
    \qquad
    D_2^{(\beta)}
    \;=\;
    \frac{\mathrm{ECF}(3,\beta)\, \mathrm{ECF}(1,\beta)^{3}}
         {\mathrm{ECF}(2,\beta)^{3}},
    \label{eq:c2d2}
\end{equation}
have been shown to be optimal observables for two-prong discrimination under
infrared- and collinear-safe (IRC) conditions~\cite{Larkoski:2014gra,Larkoski:2015kga}.

The next-order observables
\begin{equation}
    C_3^{(\beta)}
    \;=\;
    \frac{\mathrm{ECF}(4,\beta)\, \mathrm{ECF}(2,\beta)}
         {\mathrm{ECF}(3,\beta)^{2}}
    \label{eq:c3}
\end{equation}
and the generalized correlator
\begin{equation}
    N_3^{(\beta)}
    \;=\;
    \frac{{}_{2}e_{4}^{(\beta)}}{\bigl({}_{1}e_{3}^{(\beta)}\bigr)^{2}}
    \label{eq:n3}
\end{equation}
target three-prong substructure such as the full $t \to W b \to
q\bar{q}b$ decay topology~\cite{Moult:2016cvt}.

We compute these observables in a pure NumPy~\cite{harris2020array} implementation using
constituent four-momenta and conventional definition as in~\cite{Larkoski:2013eya}.

\subsection{The Particle Transformer}
\label{subsec:part}

The Particle Transformer~\cite{Qu:2022mxj} processes a jet of $P$ constituent
particles (zero-padded to a fixed maximum) using a stack of $L$
\emph{particle attention} blocks followed by two \emph{class attention}
blocks. The per-particle features are projected to a $d$-dimensional
embedding by a multi-layer perceptron, yielding an initial residual
stream $x^{0} \in \mathbb{R}^{P \times d}$. The pairwise features are
encoded by a stack of pointwise 1D convolutions (i.e., a position-wise
MLP applied independently to each particle pair $(i,j)$) with GELU
nonlinearities and batch normalization between layers, producing a
shared pairwise embedding tensor $U \in \mathbb{R}^{P \times P \times
d'}$, where $d'$ is chosen equal to the number of attention heads
$N_{\mathrm{heads}}$. The same $U$ is used as a bias by every particle
attention block.

For head $h$ in particle attention block $l$, the standard scaled
dot-product attention is augmented by the $h$-th channel of $U$ as a
pre-softmax additive bias:
\begin{equation}
    A^{(l,h)}_{ij}
    \;=\;
    \mathrm{softmax}_{j}\!\left[
        \frac{Q^{(l,h)}_i\, K^{(l,h)\,\top}_j}{\sqrt{d_h}}
        \;+\; U^{(h)}_{ij}
    \right],
    \label{eq:attention}
\end{equation}
where $d_h = d / N_{\mathrm{heads}}$ is the per-head embedding dimension
and $Q^{(l,h)}, K^{(l,h)}$ are linear projections of the layer-$l$
residual stream. The pairwise features feeding $U$ are the four quantities
\begin{align}\label{eqn:pairfeat}
    \ln k_{T,ij}, \quad \ln z_{ij}, \quad \ln \Delta_{ij}, \quad \ln m^{2}_{ij},
\end{align}
with
\begin{align}
    k_{T,ij} &= \min(p_{T,i}, p_{T,j})\, \Delta R_{ij}, \\
    z_{ij}   &= \frac{\min(p_{T,i}, p_{T,j})}{p_{T,i} + p_{T,j}}, \\
    \Delta_{ij} &= \Delta R_{ij}, \\
    m^{2}_{ij} &= (p_i + p_j)^2,
\end{align}
where the invariant mass $m^{2}_{ij}$ is Lorentz-invariant in the full
sense and the remaining three are invariant under longitudinal boosts.
The pairwise bias allows each attention head to express a learned
attention pattern that is informed by physical pairwise kinematics
beyond what is available in the per-particle features alone.

Each particle attention block follows the NormFormer
layout~\cite{2021arXiv211009456S}: layer normalization both before and after
the multi-head attention, layer normalization before each linear layer
of the feed-forward sub-block, and residual connections around each
sub-block. The output of each attention head is multiplied by a learned
scalar gain $c_{l,h}$ before being summed into the residual stream;
this is the head scaling that we exploit for zero ablation in
Section~\ref{subsec:interventions}.

The class attention blocks~\cite{2020arXiv201011929D} aggregate
the particle representations into a single classification vector via a
learnable token $x_{\mathrm{class}} \in \mathbb{R}^{d}$. In each class
attention block, the multi-head attention takes its query from
$x_{\mathrm{class}}$ alone, while the keys and values are computed from
the concatenation $[x_{\mathrm{class}}, x^{L}]$ of the class token with
the output of the final particle attention block. The two class
attention blocks act sequentially on $x_{\mathrm{class}}$; the final
class token is passed through a layer normalization and a linear
classification head producing two output logits for the top-versus-QCD
classification.